\documentstyle[prd,aps,epsf,eqsecnum,amssymb]{revtex}
\begin{document}

\title{ Second order perturbations in the radius stabilized 
\\
Randall-Sundrum two branes model} 

\author{  
    Hideaki Kudoh\footnote{ Email address: kudoh@yukawa.kyoto-u.ac.jp}
  and
   Takahiro Tanaka\footnote{ Email address: tanaka@yukawa.kyoto-u.ac.jp}
 }
 \address{Yukawa Institute for Theoretical Physics, 
          Kyoto University, Kyoto 606-8502, Japan}
\vspace{5mm}
\maketitle

\begin{abstract}
The nonlinear gravitational interaction is investigated in the
Randall-Sundrum two branes model with the radius stabilization
mechanism. As the stabilization model, we assume a single scalar
field that has a potential in the bulk and a potential on each
brane. We develop a formulation of the second order gravitational
perturbations under the assumption of a static and axial-symmetric five-dimensional metric that is spherically symmetric in the
four-dimensional sense.  After deriving the formal solutions for the
perturbations, we discuss the gravity on each brane induced by the
matter on its own side, taking the limit of large
coupling of the scalar field interaction term on the branes. 
We show using the Goldberger-Wise stabilization model that four-dimensional Einstein gravity is approximately recovered in the
second order perturbations.

\vspace*{1mm}
\noindent
PACS 04.50.+h; 98.80.Cq
\\
YITP-01-82
\end{abstract}


\newcommand{\bsA}{  {\bar{\mathcal{A}}}  }
\newcommand{\bsAf}{ {\bar{\mathcal{A}}}^{(1)}  }
\newcommand{\bsAs}{{\bar{\mathcal{A}}}^{(2)}  }
\newcommand{\bsB}{ {\bar{\mathcal{B}} }  }
\newcommand{\bsBf}{ {\bar{\mathcal{B}}^{(1)} }  }
\newcommand{\bsBs}{ {\bar{\mathcal{B}}^{(2)} }  }
\newcommand{\bsC}{ {\bar{\mathcal{C}} }  }
\newcommand{\bsCf}{ {\bar{\mathcal{C}}^{(1)} }  }
\newcommand{\bsCs}{ {\bar{\mathcal{C}}^{(2)} }  }
\newcommand{\sA}{ {\mathcal{A}}  }
\newcommand{\sAf}{ {\mathcal{A}}^{(1)}  }
\newcommand{\sAs}{ {\mathcal{A}}^{(2)}  }
\newcommand{\sB}{ {\mathcal{B}}}
\newcommand{\sBs}{ {\mathcal{B}}^{(2)}  }
\newcommand{\sC}{ {\mathcal{C}}  }
\newcommand{\sCf}{ {\mathcal{C}}^{(1)}  }
\newcommand{\sCs}{ {\mathcal{C}}^{(2)}  }
\newcommand{\sG}{ {\mathcal{G}}  }
\newcommand{\sL}{ {\mathcal{L}}  }
\newcommand{\sO}{ {\mathcal{O}}  } 
\newcommand{\sP}{ {\mathcal{P}}  } 
\newcommand{\sR}{ {\mathcal{R}}  } 
\newcommand{\sS}{ {\mathcal{S}}  } 
\newcommand{\sT}{ {\mathcal{T}}  } 

\newcommand{\br}{\bar{r}}
\newcommand{\by}{\bar{y}}
\newcommand{\bg}{\bar{g}}
\newcommand{\bw}{\bar{w}}
\newcommand{\baf}{\bar{f}}
\newcommand{\stac}[2]{ \stackrel{\scriptscriptstyle {#1}}{#2}   }



   \section{Introduction}

Many unified theories require spacetime dimensions higher than that observed in the Universe, and thus the extra dimensions must be invisible by some mechanism. One of the possible schemes is known as Kaluza-Klein compactification. Recently, theories with extra dimensions have attracted considerable attention from the other viewpoint of providing a solution to the hierarchy problem
\cite{Arkani-Hamed:1998rs,Randall:1999ee,Randall:1999vf}
\cite{Lykken:2000nb,Mukohyama:2001wq}. 
The main idea to resolve the large hierarchy is that the small coupling of four-dimensional gravity is generated by the large physical volume of extra dimensions. 
These theories provide a novel setting for discussing phenomenological, cosmological, and conceptual issues that are related to extra dimensions.

The model that was introduced by Randall and Sundrum (RS) is particularly attractive. The RS two branes model is constructed in a five-dimensional anti-de Sitter (AdS) spacetime \cite{Randall:1999ee}.
 The fifth coordinate is compactified on $S^1/Z_2$, and the positive and negative tension branes are on the two fixed points. 
It is assumed that all matter fields are confined on each brane and only the gravity propagates freely in the five-dimensional bulk.
In this model, the hierarchy problem is resolved on the brane with negative tension if the separation of the branes is about 37 times the AdS radius.

Apart from the fine tuning of the brane tension that is necessary to solve the cosmological constant problem, one of the significant points in discussing the consistency of this model is whether four-dimensional Einstein gravity is recovered on the brane
\cite{Garriga:2000yh,Tanaka:2000er,Shiromizu:2000wj,Sasaki:2000mi,Giddings:2000mu,Tanaka:2000zv,Mukohyama:2001ks}.
Another point is to give a so-called radius stabilization mechanism that works to select the required separation distance between the two branes to resolve the hierarchy without fine-tuning \cite{Goldberger:1999uk,Goldberger:2000un,DeWolfe:2000cp,Goldberger:2000dv,Arkani-Hamed:2001kx,Luty:2001ec,Hofmann:2001cj,Garriga:2001jb,Flachi:2001bj,Nojiri:2000bz,Brevik:2001vt}.  
The stabilization mechanism not only is important to guarantee a stable hierarchy, but also plays an important role in the recovery of four-dimensional Einstein gravity in linear order \cite{Tanaka:2000er,Mukohyama:2001ks} and of the correct cosmological expansion \cite{Csaki:1999jh,Binetruy:2000hy,Flanagan:2000dc,Csaki:2000mp,Kim:2000hi}.
The discussion of the recovery is almost independent of the details of the stabilization model. 
The essence in recovering the four-dimensional Einstein gravity in linear perturbations is that the massless mode of the scalar-type gravitational perturbation disappears due to the bulk scalar field, and only the tensor-type perturbation continues to have a massless mode. 
When the stabilization mechanism is turned off, the induced gravity on the brane becomes of the Brans-Dicke type with an unacceptable Brans-Dicke parameter \cite{Garriga:2000yh}.

A large number of studies have been made of the gravity in the brane world model \cite{Chamblin:2000cj,Chamblin:2000by,Emparan:2000wa,Charmousis:2000rg,Gen:2001nu,Germani:2001du,Wiseman:2001xt,Lewandowski:2001qp,Mukohyama:2001jv}.
 Although the model does not have a drawback in linear perturbation, it is not a trivial question whether the second order gravitational perturbation works as well. For the second order perturbation in the RS single brane model without bulk scalar field, where the tension of the brane is positive, it has been confirmed that there is no observable disagreement with four-dimensional Einstein gravity \cite{Kudoh:2001wb,Giannakis:2001zx}. However the setting of the RS two brane model with stabilization mechanism is quite different from that of the single brane model, and furthermore we are mainly concerned with the gravity on the negative tension brane. In this paper we study the second order gravitational perturbation of the
RS two branes model with a stabilization mechanism due to a bulk scalar field. 
 To simplify the analysis, we consider static and axisymmetric configurations, which means that the metric on the branes is spherically symmetric. After developing a formulation to calculate the second order perturbation, we take the limit that the coupling of the scalar field interaction term on each brane is very large. In this limit, we find that four-dimensional Einstein gravity is approximately recovered.

The paper is organized as follows. In Sec. \ref{sec:Perturbation
Equations} we describe the model that we will study, and derive the second order perturbation equations in the five-dimensional bulk. We also discuss the gauge transformations and the boundary conditions. In Sec. \ref{sec:mode} we explain our approximation scheme, and give the formal solutions. 
In Sec. \ref{sec:4D recovery-1st} we review the results for linear perturbations, giving their explicit expressions following the notation of this paper, and explain the setup of the problem that we study in the present paper. In Sec. \ref{sec:4D recovery-2nd} we analyze the second order metric perturbations induced on each brane. We show that the four-dimensional Einstein gravity is recovered with some small corrections. 
These results are summarized in Sec.  \ref{sec:summary}.

   \section{Perturbation Equations in the RS model}
   \label{sec:Perturbation Equations}

We consider the second order perturbations in the RS two branes model with a five-dimensional scalar field introduced to stabilize the distance between the two branes. 
According to the warped compactification of the RS model, the unperturbed metric is supposed to be 
\begin{eqnarray}
    ds^2 = g_{ab}dx^a dx^b 
     =dy^2 +a^2(y) \eta_{\mu\nu} dx^\mu dx^\nu  \,,
\end{eqnarray}
where $\eta_{\mu\nu}$ is the four-dimensional Minkowski metric with $(-\,+\,+\,+)$ signature. 
The $y$ direction is bounded by two branes located at $y=y_{+}$ and
$y=y_{-}$, whose tensions are assumed to be positive ($\Lambda_{(+)}
>0$) and negative ($\Lambda_{(-)} <0$), respectively. On these two
branes, $Z_2$ symmetry is imposed, and we adopt the convention
$y_{+}<y_{-}$.
To generate the hierarchy between Planck and electroweak scales, we need 
\begin{eqnarray}
    \frac{ a_+ }{ a_- } 
    \sim 10^{16} \,,
\label{eq:a+/a_}
\end{eqnarray}
where $a_\pm \equiv a(y_\pm)$.

In this paper we investigate the gravity induced by non-relativistic matter fields confined on each brane whose energy-momentum tensor is given in the perfect fluid form: 
\begin{eqnarray}
    T_{{\pm} \nu }^{ ~~ \mu} = 
    a_\pm^{-4} \,
    {\mathrm{diag}} 
    \{- \rho_{\pm}, P_{\pm}, P_{\pm}, P_{\pm} \}.
\label{eq:EM tensor}
\end{eqnarray}
The warp factor in the definition of the energy-momentum tensor~(\ref{eq:EM tensor}) is incorporated for the following reason. In the present analysis, we adopt the normalization that any physical quantities are always mapped onto and measured by the length scale at $y=y_ +$. Since the length scale is warped by a warp factor $a(y)$,  physical quantities such as $\rho_-$ and $P_-$ are scaled by a factor $a_-^{-4}$. 

 To simplify the analysis, we restrict our consideration to a static and axisymmetric spacetime whose axis of symmetry lies along $y$ direction. 
We denote the perturbed metric by $\tilde{g}_{ab}=g_{ab}+h_{ab}$. 
The four-dimensional perturbation $h_{\mu\nu}$ is divided into a trace part and a transverse-traceless (TT) part. According to this decomposition, we assume that the perturbed metric has the diagonal form 
\begin{eqnarray}
 ds^2 
      &=& e^{2Y(r,y)} dy^2 + a^2(y) 
\left[- e^{A(r,y)- \psi(r,y)} dt^2
      + e^{B(r,y)- \psi(r,y)} dr^2
      + e^{C(r,y)- \psi(r,y)} r^2 d\Omega^2 
\right].
\label{eq:metric}
\end{eqnarray}
Here $A$, $B$, and $C$ correspond to the TT part, and $\psi$ to the trace part.  The TT condition to the linear order is given in terms of $A, B$, and $C$ as
\begin{eqnarray}
   A^{(1)}(r,y)&=&-\frac{1}{r^2}\partial_r (r^3 B^{(1)}(r,y))\,,
   \nonumber
\\
   C^{(1)}(r,y)&=& \frac{1}{2r }\partial_r (r^2 B^{(1)}(r,y))\,,
    \label{eq:TT cond}
\end{eqnarray}
where we have expanded the metric functions up to second order as  
\begin{eqnarray}
    A=\sum_{J=1,2} A^{(J)}.
\end{eqnarray}
The other metric functions $Y$ and $\psi$ are expanded in the same way.  Henceforth we neglect higher order terms without mentioning it, and we omit the superscript indicating the order when it is obvious.  
We impose the same conditions as Eq.~(\ref{eq:TT cond}) on $A^{(2)},  B^{(2)}$, and $C^{(2)}$ so that $B^{(J)}$ and $C^{(J)}$ are derived from $A^{(J)}$ once it is solved. Since the trace of the metric is given by 
\begin{eqnarray}
   \tilde g^{\mu \nu}h_{\mu \nu} 
   &=& -4\psi - 2\psi^2- \frac{1}{2} (A^2+B^2+2 C^2)   \,,
\end{eqnarray}  
the trace part at linear order is $\psi^{(1)}$, while 
$\psi^{(2)}$ does not correspond to the trace part at second order.
Hence the second order counterpart of the condition (\ref{eq:TT
cond})  does not mean that $A^{(2)}$ is the transverse-traceless perturbation, but we extensively refer to these metric functions as the TT part and $\psi$ as the trace part. 
Later we will show that $Y$ coincides with $\psi$ in a linear perturbation, and thus this metric assumption is the same at least in linear perturbation as the ``Newton gauge" condition.

The Lagrangian for the bulk scalar field is  
\begin{equation}
 {\cal L}= - {1\over 2} \tilde  g^{ab} 
     \tilde\varphi_{,a} \tilde\varphi_{,b}
          -V_B(\tilde\varphi)
          -\sum_{\sigma=\pm} V_{(\sigma)}(\tilde\varphi)
              \delta(y-y_{\sigma}) \,,
\label{eq:Lagrangian}
\end{equation}
where $V_B$ and $V_{(\pm)}$ are the potential in the bulk and that on the corresponding brane.
For most of the present analysis, we do not need to specify the explicit form of the potentials 
$V_B$ and $V_{(\pm)}$. 
The scalar field is expanded up to second order as 
\begin{eqnarray}
 \tilde\varphi (r, y) 
 = \phi_0(y) + \varphi^{(1)}(r,y)+\varphi^{(2)}(r,y) \,,
\end{eqnarray}
where $\phi_0$ is the background scalar field configuration, which depends only on $y$.

From the five-dimensional Einstein equations with the cosmological term $\Lambda$ and the equation of motion for the scalar field, we obtain the background equations as
\begin{eqnarray}
 \dot H(y) &=& -{\kappa\over 3}\dot\phi_0^2(y),
\cr
    H^2(y) &=& {\kappa \over 6}
    \Bigl( {1\over 2} \dot\phi_0^2(y) 
        - V_B(\phi_0(y))-\kappa^{-1}\Lambda  
    \Bigr),
\cr 
 \ddot\phi_0(y) &+& 4 H(y) \dot \phi_0(y) -V_B'(\phi_0(y))=0,
\label{eq:background eqs}
\end{eqnarray}
where $H(y):= \dot a(y)/a(y) \bigl(\approx - \sqrt{-\Lambda/6}\bigr)$
and the five-dimensional Newton's constant is $G_5=\kappa/8\pi$.  
An overdot denotes differentiation with respect to $y$.

    \subsection{5D Einstein equations in the bulk}

In this subsection we derive the master equations for $A^{(J)}$ and  $Y^{(J)}$ from the five-dimensional Einstein equations. 
From the $(r,r)$, $(\theta, \theta)$, and $(r,y)$ components of the Einstein equations, we obtain two independent equations:
\begin{eqnarray}
  &&  \partial_r [ \Delta (\psi^{(J)} - Y^{(J)} )]
    = \epsilon^{(J)} \partial_r S_{\psi} \,,
\nonumber
\\ 
   &&   \varphi^{(J)} 
   =  \frac{3}{2\kappa \dot \phi_0}
     ( \partial_y \psi^{(J)} + 2H Y^{(J)} 
   + \epsilon^{(J)} S_{\varphi}) \,,
\end{eqnarray}
where $\Delta \equiv \sum_{i=1}^3 \partial_i^2$. Here 
 $S_{\psi}$ and $S_{\varphi}$ are the second order source terms that are constructed from the first order quantities, and the explicit forms are given below.
We have introduced the symbol $\epsilon^{(J)}$ that is defined by $\epsilon^{(1)}=0$ and $\epsilon^{(2)}=1$ to represent the first and the second order equations in a single expression. These equations are reduced to 
\begin{eqnarray}
   && \psi^{(J)} (r,y) = Y^{(J)} + \epsilon^{(J)} \Delta^{-1} S_\psi  \,,
\label{eq: phi^J=Y^J+...}
\\
   &&   \varphi^{(J)} (r,y)
    = \frac{3}{2\kappa \dot \phi_0 a^2}  \partial_y( a^2 Y^{(J)})
    + \frac{3}{2\kappa \dot \phi_0} \epsilon^{(J)} \Bigl[
      S_\varphi 
    + \partial_y \Delta^{-1} 
  S_\psi 
    \Bigr] \,.
\label{eq:d varphi = Y^J +...}
\end{eqnarray}

The equations for $A^{(J)}$ and $Y^{(J)}$ are obtained from the $(r,r)$ and $(y,y)$ components of Einstein equations as
\begin{eqnarray}
   &&  [ \hat L^{(TT)}+\frac{1}{a^2} \Delta ] (a^2A^{(J)})= \epsilon^{(J)} S_A \,,
\label{eq:Ein-A^J}   
\\
   &&  [\hat L^{(Y)}+ \Delta ] Y^{(J)}= \epsilon^{(J)} S_Y\,,
\label{eq:Ein-Y^J}
\end{eqnarray}
where 
\begin{eqnarray*}
 &&  \hat L^{(TT)} 
 :={1\over a^2}\partial_y a^4\partial_y {1\over a^2} \,, 
\\
  && \hat L^{(Y)}
  := a^2\dot\phi_0^2\partial_y{1\over a^{2}\dot\phi_0^2}
      \partial_y a^2-{2\kappa\over 3}a^2\dot\phi_0^2 \,.
\end{eqnarray*}
The second order source terms $S_A$ and  $S_Y$ are given later. 
After the simplification using the linear order equations (\ref{eq:
phi^J=Y^J+...}), (\ref{eq:d varphi = Y^J +...}), (\ref{eq:Ein-A^J}), and  (\ref{eq:Ein-Y^J}), $S_{\psi}$ and $S_{\varphi}$ are written down
as
\begin{eqnarray}
   \partial_r S_\psi (r,y) & = &
 \frac{3}{8r^6}\partial_r (r^8 (\Delta B )^2)
    + B_{,r} 
  \Bigl(
      \frac{14}{3} \Delta B 
   + \frac{5}{4}r \partial_r \Delta B 
   - \frac{11}{6r} B_{,r}
  \Bigr)
+ 
 \frac{9}{ 4r^{8/3}} \partial_r
 \Bigl\{  r^{8/3} 
  \frac{ Y_{,r}(\Delta Y)_{,r}}{ \kappa a^2 \dot\phi_0^2} 
 \Bigr\} 
\nonumber
\\
&&  + 
  \frac{1}{a^2} \partial_y
 \biggl(
\underline{ 
  - 2 a^4 Y_{,r} B_{,y} 
  + \frac{3}{2r^{8/3}} \partial_r
    \Bigl\{ r^{8/3} 
  \frac{ a^2 Y_{,r} \varphi_{,r}}{\dot\phi_0} 
  \Bigr\} 
}
 \biggr) \,,
\label{eq:Def. S_Psi}
\end{eqnarray}   
and 
\begin{eqnarray}
   S_\varphi 
&=& {\mathbb S}_\varphi 
  -  \frac{ \varphi }{ a^2 \dot\phi_0}
   \Bigl( 
    \Delta Y 
  - \frac{\kappa a^2 \ddot\phi_0}{3}  \varphi  
  \Bigr)\,,
  \label{eq:Def. S_varphi}
\end{eqnarray}
where 
\begin{eqnarray}
{\mathbb S}_\varphi  &=& \frac{2}{3} \int 
\Bigl(  B_{,y} A_{,r}  
     + \frac{r}{8}  
    B_{,y r}
        (3A_{,r}+B_{,r})
    \Bigr)
    dr
     -   \int B _{,y} Y _{,r}  
    dr
  +  \frac{1}{ a^2  \dot\phi_0 }
     \int  \varphi_{,r}  \Delta Y dr
  +  H Y^2
  -  \frac{\kappa}{3} \partial_y       
     \bigl( \varphi^2 \bigr) \,.
  \label{eq:Def. mathbb S_varphi}
\end{eqnarray}
In the derivation of Eq.~(\ref{eq:Def. S_Psi}), we  used Eq.~(\ref{eq:Y^2+varphi^2}).  The underline in Eq.~(\ref{eq:Def. S_Psi}) is attached for convenience of our explanation. So is the double underline below.
Using Eq.~(\ref{eq:Def. S_Psi}),  $S_A$ is given by 
\begin{eqnarray}
  S_A 
  &=&
  -\frac{8}{r} B B _{,r}
    -\frac{17}{3}( B _{,r})^2
  - \frac{1}{r^6}\partial_r(r^7 B B _{,rr})
  - \int dr
  \Bigl(   \frac{r}{2} (B _{,rr})^2
         + \frac{7}{r} (B _{,r})^2
  \Bigr)
\nonumber
\\ 
& &
-  3 \int \frac{dr}{r} 
  \Bigl(
  \frac{ Y_{,r}(\Delta Y)_{,r}}{ \kappa a^2 \dot\phi_0^2}     \Bigr)
+  \frac{1}{a^2} \partial_y
   \biggl[ 
   \underline{\underline{  3a^4 Y A_{,y}}}
 + \underline{  \int dr (a^4 Y_{,r} B_{,y})
    -  2 \int \frac{dr}{r}
       \frac{a^2 Y_{,r}\varphi _{,r}}{\dot\phi_0} 
   }
 \biggr] 
 \,,
\label{eq:S_A full exp}
\end{eqnarray}
where we  again used Eq.~(\ref{eq:Y^2+varphi^2}). 
The complete expression for $S_Y$ is slightly complicated. 
For later convenience we divide $S_Y$ as 
\begin{eqnarray}
    S_Y 
&=& 
{\mathbb S}_Y 
   - {\dot\phi_0} ^2 a^2
    \partial_y 
    \left(  \frac{1}{\dot \phi_0^2} 
    \Bigl[
       {\mathbb S}_\varphi 
    +  \partial_y \Delta^{-1} S_\psi 
    \Bigr]
    \right) \,,
\label{eq:S_Y3+...}
\end{eqnarray}
where  ${\mathbb S}_Y$ is given by 
\begin{eqnarray}
{\mathbb S}_Y
 &=& 
 - \frac{a^2 }{8}
     \Bigl( (A _{,y})^2 + (B_{,y})^2 
          + \frac{2}{3} A _{,y} B_{,y} 
     \Bigr)
 - \int \biggl\{ \Delta B  
      \Bigl(  \frac{ 7 r }{3} \Delta B 
   + \frac{r^2}{2} (\Delta B)_{,r}
   + 4 B _{,r} 
     \Bigr)
  + \frac{5 r }{6} B_{,r}
 (\Delta B )_{,r}
  +  \frac{(B _{,r})^2}{3 r} 
\cr 
&&
  + \frac{a^2 H}{3}
\Bigl[
    A_{,r} (3A+B)_{,y} -r B_{,r} B_{,ry}
\Bigr]
  + \Bigl[
    Y_{,r} \Bigl( \Delta B + \frac{8 B_{,r}}{r} \Bigr)
  - B (\Delta Y )_{,r}
  - 2B_{,r}  \Delta Y 
    \Bigr]
  + 3 Y_{,r} \Delta Y   
\cr
&&
+ \Bigl( Y+ \frac{4H\varphi}{\dot \phi_0} \Bigr) (\Delta Y)_{,r} 
  +  \frac{2 \kappa}{3} 
    \bigl(
       \varphi_{,r} \Delta \varphi 
       - \varphi (\Delta \varphi)_{,r} 
    \bigr)
  - 2a^4 B_{,y} 
  \Bigl( \frac{Y_{,r}}{a^2} \Bigr)_{,y}
      \biggr\}  dr 
  - \frac{9 (\Delta Y )^2}{4 \kappa a^2 {\dot\phi_0 }^2}  
  - \underline{ a^2 \dot H  Y^2 } \,.
\end{eqnarray}

     \subsection{Boundary condition}
 
In the previous subsection, we derived master equations for the metric functions in the bulk up to second order. To solve these equations we must determine the boundary conditions on the branes. It is well known that the boundary condition is given by Israel's junction condition \cite{Israel:1966rt}, which is easily obtained in Gaussian normal coordinates. On the other hand, the Newton gauge simplifies the master equations for the perturbations. Therefore we consider the gauge transformation between them.

In Gaussian normal coordinates $\bar x^a$, the metric becomes
\begin{eqnarray}
 ds^2 = d\bar{y}^2 + a^2(\bar{y}) 
\left[- e^{\bar{A} } d\bar{t}^2
      + e^{\bar{B} } d\bar{r}^2
      + e^{\bar{C} } \bar{r}^2 d \Omega^2 
\right] \,,
\label{eq:GN coordinate}
\end{eqnarray}
with $\bar y= \text{const}$ on either the positive or negative tension brane. Note that we introduced two sets of Gaussian normal coordinates; one is the coordinate set in which the positive tension brane is located at $\bar y= \bar y_+$, and the other is that in which the negative tension brane is located at $\bar y= \bar y_-$.
Corresponding to these two different Gaussian normal coordinates, there are two infinitesimal gauge transformations $\bar{x}^a=x^a+\xi_\pm^a(x)$ between the Newton gauge and the Gaussian normal gauge, respectively. 
To satisfy the restriction on the metric form in both gauges, the gauge parameters $\xi_{\pm}^a= \stac{(1)}{\xi_{\pm}^a} + \stac{(2)}{\xi_{\pm}^a}$ associated with each brane must take the form of 
\begin{eqnarray}
 \stac{(J)}{\xi ^{y}_{\pm}} (r,y)
 &=& \int_{y_{\pm}}^y dy' 
 \left( 
    Y^{(J)} +\frac{1}{2} \epsilon^{(J)}
 \Bigl[ Y ^2 - \frac{1}{a^2} (\stac{(1)}{\xi^y_{,r}})^2
 \Bigr]
 \right)
    + \stac{(J)}{\hat\xi^y_{\pm}}(r), 
\label{eq:xi^y}
\\
    \stac{(J)}{\xi ^{r}_{\pm} }(r,y)
    &=& \int^y_{y_{\pm}} dy'  
    \left( 
    - \frac{1}{a^2}   \stac{(J)~~~}{\xi^{y}_{\pm,r}} 
    +\frac{\epsilon^{(J)}}{a^2} \stac{(1)~~~}{\xi^{y}_{\pm,r}} 
    \Bigl[
      \bar B - Y +2 H \stac{(1) }{\xi^{y}_{\pm}} 
       + \stac{(1)~~~}{\xi^{r}_{\pm,r}} 
    \Bigr]
    \right)
+ \stac{(J)}{\hat\xi^r_{\pm}}(r)  ,
\label{eq:xi^r}
\end{eqnarray}
where we simplified the integrand of the equation for the second order perturbation by using the result for a linear perturbation. 
The functions of $r$, ${\hat\xi^r_{\pm}}$ and ${\hat\xi^y_{\pm}}$, arise as integration constants. The arbitrariness of ${\hat\xi^r_{\pm}}$ is due to a residual gauge degree of freedom of the coordinate transformation in the radial direction, while ${\hat\xi^y_{\pm}}$ is determined with the aid of the junction conditions as we will see below.

The gauge transformations for each metric component are given by 
\begin{eqnarray}
  \bar{A}^{(J)} (r,y) &=& A^{(J)}-\psi^{(J)}-2H \stac{(J)}{\xi_{\pm}^{y}}   
  - \epsilon^{(J)} 
  \Bigl[
     \stac{(1)}{\xi_{\pm}^y} \bar{A}^{(1)}_{,y} + \stac{(1)}{\xi_{\pm}^r} \bar{A}^{(1)}_{,r} + \dot H ( \stac{(1)}{\xi_{\pm}^y})^2   
  \Bigr] \,,
\nonumber
\\
   \bar{B}^{(J)} (r,y) &=& B^{(J)}-\psi^{(J)}-2H\stac{(J)}{\xi_{\pm}^{y}} 
   - 2\stac{(J)}{\xi_{\pm,r}^{r} } 
   - \epsilon^{(J)} 
  \Bigl[
     \stac{(1)}{\xi_{\pm}^y} \bar{B}^{(1)}_{,y} 
   + \stac{(1)}{\xi_{\pm}^r} \bar{B}^{(1)}_{,r} 
   - ( \stac{(1)~~}{\xi_{\pm ,r }^r})^2   
  + \frac{1}{a^2} ( \stac{(1)~~}{\xi_{ \pm ,r }^y})^2  
   + \dot H ( \stac{(1)}{\xi_{\pm}^y})^2   
  \Bigr] \,,
\nonumber
\\
   \bar{C}^{(J)} (r,y)&=&C^{(J)}-\psi^{(J)}-2H\stac{(J)}{\xi_{\pm}^{y}} 
   - \frac{2}{r} \stac{(J)}{\xi_{\pm}^{r} } 
   - \epsilon^{(J)} 
  \Bigl[
     \stac{(1)}{\xi_{\pm}^y} \bar{C}^{(1)}_{,y} 
   + \stac{(1)}{\xi_{\pm}^r} \bar{C}^{(1)}_{,r} 
   - \frac{1}{r^2} ( \stac{(1)}{\xi_{\pm}^r})^2  
   + \dot H ( \stac{(1)}{\xi_{\pm}^y})^2   
  \Bigr] \,.
\label{eq:gauge trans ABC}
\end{eqnarray}
As for the scalar field, its gauge transformation is given by 
\begin{eqnarray}
   \bar{\phi}_0(y)  &=&  \phi_0(y) \,,
\nonumber
 \\
   \bar{\varphi}^{(J)} (r, y) 
   &=& \varphi^{(J)}(r,y) - \delta \varphi^{(J)}(r,y) \,,
\label{eq:bar varphi^J to varphi^J}
\end{eqnarray}
where 
\begin{eqnarray}
  \delta \varphi^{(J)}(r,y) 
  &=&  
    \dot{\phi_0} \stac{(J)}{\xi_{\pm}^y}
   + \epsilon^{(J)} 
   \left( 
     \stac{(1)}{\xi_{\pm}^y} { {\varphi}}^{(1)}_{,y}
   + \stac{(1)}{\xi_{\pm}^r} 
   ({ {\varphi}}^{(1)} - \dot\phi_0 \stac{(1)}{\xi^y_{\pm}})_{,r}
 - \frac{1}{2} 
   \Bigl[ \dot\phi_0 (\stac{(1)}{\xi_{\pm}^y})^2 
   \Bigr] _{,y}
   \right)\,.
\end{eqnarray}

As mentioned earlier, we assume the energy-momentum tensor to be of the perfect fluid form (\ref{eq:EM tensor}). 
The four-dimensional energy-momentum conservation $T^{~\nu \mu}_{ \pm ~;\nu}=0$ becomes 
\begin{eqnarray}
 (\rho_{\pm} +P_{\pm}) \partial_r {\bar{A}}^{(1)}(r,y_{\pm}) 
       + 2 \, \partial_r P_{\pm} =0  \,,
\label{eq:energy momentum cons}
\end{eqnarray}
and hence we find that $P_{\pm}$ is a second order quantity. This equation represents the force balance between pressure and gravity acting on the matter field.

Now we consider the boundary conditions. Israel's junction conditions on the 3-branes are given by
\begin{eqnarray}
   \pm  \tilde{ \bar{g}} ^{\nu \lambda} 
   \partial_{ y} ( \tilde{ \bar{g}}_{\mu \lambda} ) 
  = -\kappa  \Bigl[T^{~\nu}_{ \mu} -{1\over 3} \delta_\mu^\nu T \Bigr]_\pm 
  -\kappa  \Bigl[ T^{(\varphi)\nu}_{ \mu} -{1\over 3}\delta_\mu^\nu T^{(\varphi) } \Bigr]_\pm
     - \frac{\kappa }{3} \delta^\mu_\nu \Lambda_{(\pm)} \, 
    ~~ ({\mathrm{at}}~ y=y_{\pm})   \,,
\label{eq:Israel junc}
\end{eqnarray}
where 
$\Lambda_{(\pm)}$ is the tension on each brane, and $T^{(\varphi)}_{\mu\nu}$ is the energy-momentum tensor for the scalar field. Here and hereafter, when we evaluate the value at $y=y_\pm$, we take the value at $y=y_\pm\pm\varepsilon$ in the $\epsilon\to 0$ limit. 
Since, by assumption, the scalar field dose not have a kinetic term on the brane, its energy-momentum tensor on the brane is given by $T^{(\varphi) \nu}_{~~\mu} = - V_{(\pm)}\delta^\nu _\mu $. 
Then, Eq.~(\ref{eq:Israel junc}) gives at the lowest order 
\begin{eqnarray}
   \mp H  =   \frac{\kappa}{6}  V_{(\pm)} 
   + \frac{\kappa}{6} \Lambda_{(\pm)} \, 
    \quad (y=y_{\pm})\,.
\end{eqnarray}
The potential $V_{(\pm)}$ must be chosen to satisfy this condition. 
The $(t,t)$ component of the junction condition (\ref{eq:Israel junc}) gives 
\begin{eqnarray}
   \pm \partial_{ {y}} \bar{A}^{(J)} 
   = -\kappa 
    \Bigl( T^0_0-\frac{1}{3}T \Bigr)^{(J)}_{\pm}
   \mp   \frac{2 \kappa}{3}    
   \Bigl( \bar\varphi^{(J)} \dot{\bar\phi_0} 
   + \frac{\epsilon^{(J)}}{4} \partial_y[( {\bar\varphi}^{(1)})^2]
   \Bigr) 
   \quad ( y=  y_{\pm})\,.
\label{eq:jun cond. GN}
\end{eqnarray}
The junction condition in the Newton gauge is obtained by applying the gauge transformation to this equation. We define $\Sigma^{(J)}_{\pm}$ as the jump of the derivative of a metric function in the Newton gauge:
\begin{eqnarray}
   \frac{\kappa}{a_{\pm}^2} \Sigma^{(J)}_{\pm} 
   &\equiv& 
   \pm {A}^{(J)}_{,y}|_{y=y_{\pm} },  
\nonumber
\\
   &=& -\kappa \Bigl( T^0_0-\frac{1}{3}T \Bigr)^{(J)}_{ \pm }
   \mp \epsilon ^{(J)}  S_{\Sigma}^{\pm}  \, 
   \quad ( y=y_{\pm}) \,,
\label{eq:jun. for A_,y}
\end{eqnarray}
where $S_{\Sigma}^{\pm}$ is given by the substitution of
Eqs.~(\ref{eq:background eqs}), (\ref{eq:Ein-Y^J}), 
(\ref{eq:Def. S_varphi}), and (\ref{eq:bar varphi^J to varphi^J}) 
as
\begin{eqnarray}
  S_{\Sigma}^{ \pm }
   &=&  
    \frac{2 \kappa}{3} 
\Bigl( \bar\varphi^{(2)} \dot{\bar\phi_0} 
   + \frac{1}{4} \partial_y[( {\bar\varphi}^{(1)})^2]
  \Bigr)
   - ( A^{(2)}_{,y} - \bar A^{(2)}_{,y}) \, 
   \qquad (y=y_\pm)
   \,,
\nonumber
\\
&=&
 \frac{2}{3}  \int 
    \Bigl\{ 
    B_{,y} \Bigl( A_{,r} - \frac{3}{2}Y_{,r} \Bigr) 
 + \frac{r B_{,y r}}{8} (3A_{,r}+B_{,r})
 + \frac{3 \hat \xi^y_{\pm,r}  \Delta Y }{2a^2_\pm} 
   \Bigr\} dr 
-  A_{,y}(Y-4H \hat\xi^y_\pm )- \hat\xi^r_\pm A_{,ry}
\nonumber
\\
&&
+ \frac{1}{a^2_\pm} 
 \biggl( 
  {\hat\xi^y_\pm} \Delta A 
 - H (\hat\xi^y_{\pm,r})^2
+  {\hat\xi^y_{\pm,r}}  (A-Y)_{,r}
\biggr)
- 
\int dr  
  \frac{ (\varphi-\dot\phi_0 \hat\xi^y_\pm)^2}{ 2a_\pm^2 \dot\phi_0} 
  \partial_r \biggl(
  \frac{ \Delta Y }{ \varphi-\dot\phi_0 \hat \xi^y_\pm}
   \biggr) \,.  
\label{eq:S_Sigma}
\end{eqnarray}
Taking the trace of the junction condition and using Eq.  (\ref{eq:d varphi = Y^J +...}) and the formulas for the gauge transformation, we obtain the equation which determines $\hat\xi^{y}_{\pm}$:
\begin{eqnarray}
   {1 \over a_{\pm}^2} \Delta 
   \stac{(J)}{\hat\xi_{\pm}^y}
   = \pm{ \kappa\over 6} T^{(J)}_{\pm} 
   + \epsilon^{(J)} S_{\xi}^{\pm} ,
\label{eq: Lap xi^y = EM}
\end{eqnarray}
where we introduced
\begin{eqnarray}
 S_\xi^\pm &=& - \frac{4}{3} \int \Bigl(
    B_{,y} \Bigl( A_{,r} - \frac{3}{2} Y_{,r} \Bigr)
   + \frac{r}{8}   B_{,y r} (3A_{,r}+B_{,r})
   + \frac{ 3 \hat\xi^y_{\pm,r}  \Delta Y }{2a_\pm^2} 
   \Bigr) dr + \frac{1}{a_{\pm}^2} 
  \biggl( 
   {\hat\xi_{\pm,r}^y} \partial_r
     \bigl( B+Y+ H \hat\xi_{\pm}^y  \bigr)
\cr
&&
+   \bigl( B -Y - 2H {\hat\xi_{\pm}^y}
    + \hat\xi_{\pm}^r \partial_r 
   \bigr) 
   \Delta  {\hat\xi_{\pm}^y} 
  + \hat\xi_{\pm}^y \Delta Y 
  \biggr)
  + \int dr  \frac{  (\varphi-\dot\phi_0 \hat\xi^y_\pm)^2}{ a_\pm^2 \dot\phi_0} 
  \partial_r \biggl(
  \frac{ \Delta Y }{ \varphi-\dot\phi_0 \hat \xi^y_\pm}
   \biggr) \,.
\label{eq:S_xi}
\end{eqnarray}

Let us consider the junction conditions for the scalar field.
Integrating the equations of motion for the scalar field across the branes, we obtain the junction conditions for the scalar field as 
\begin{eqnarray}
    \pm 2 \dot{ \bar \phi_0 } &=& V'_{(\pm)} (\bar\phi_0) \,,
\\
    \pm 2 \dot{ \bar \varphi} ^{(J)}  
    &=& \bar{\varphi}^{(J)}V''_{(\pm)} (\bar\phi_0) 
  + \frac{1}{2}\epsilon^{(J)} 
  \Bigl[ (\bar{\varphi}^{(1)}) ^2 V'''_{(\pm)} (\bar\phi_0)  \Bigr] 
\nonumber
\end{eqnarray}
at $y=y_\pm$.
By using Eqs.~(\ref{eq:d varphi = Y^J +...}), (\ref{eq:Ein-Y^J}), 
and (\ref{eq:bar varphi^J to varphi^J}), 
the junction conditions for the scalar field in the Newton gauge 
is obtained as 
\begin{eqnarray}
  \pm 2 \dot{ \phi_0} &=&  V'_{(\pm)}( \phi_0) \,,
\nonumber
\\ 
 \frac{2}{\lambda_{\pm}}( \varphi^{(J)} - \dot\phi_0   
 \stac{(J)~~}{\hat\xi_{\pm}^y)}
 &=& 
  \mp \frac{3}{ \kappa a^2\dot\phi_0}
     \Delta Y^{(J)}
   + \epsilon^{(J)}  S^{ \pm }_{jun},  
\label{eq:Jun of varphi}
\end{eqnarray}
where we have defined 
\begin{equation}
   \lambda_{\pm}
   :={2\over V''_{(\pm)}\mp 2(\ddot\phi_0/\dot\phi_0)}  
\end{equation}
and 
\begin{eqnarray}
  S_{jun}^{\pm}
   &=&
 ( \delta \varphi^{(2)} - \dot\phi_0 \stac{(2)}{\xi^y} )
   V''_{(\pm)}
 - \frac{1}{2} (\varphi^{(1)}- \dot\phi_0 \stac{(1)}{\xi^y} )^2  
   V'''_{(\pm)}
  \pm 2 
\biggl[
  -\partial_y (\delta \varphi^{(2)}-\dot\phi_0 \stac{(2)}{\xi^y})
 - \frac{\dot\phi_0 }{2} 
   \Bigl( Y^2-\frac{ (\xi^y_{,r})^2}{a^2}  \Bigr)
\nonumber
\\
 && + \frac{3}{2\kappa a^2 \dot\phi_0} {\mathbb S}_Y
    + \dot\phi_0 \partial_y 
     \Bigl(
   \frac{ \ddot\phi_0  \varphi ^2}{2 \dot\phi_0 ^3} 
  - \frac{ 3\varphi \Delta Y}{ 2\kappa a^2 \dot\phi_0^3}
     \Bigr) 
\biggr]  
\qquad (y=y_{\pm}) \,.
\label{eq:def. S_jun}
\end{eqnarray}
A more explicit expression is given in Appendix \ref{appen:equations}.

Incorporating the boundary conditions (\ref{eq:jun. for A_,y}) and
(\ref{eq:Jun of varphi}), the master equations become  
\begin{eqnarray}
    [ L^{(TT)}+\frac{1}{a^2} \Delta ] (a^2A^{(J)})
    &=& 2\kappa \sum_{\sigma=\pm} \Sigma^{(J)}_{\sigma}
    \delta(y-y_{\sigma}) + \epsilon^{(J)} S_A  \,,
\label{eq:A^J with Junc}
\\
\label{eq:Y^J with Junc}
  [\hat L^{(Y)} + \Delta] Y^{(J)} 
  &=& 
      \sum_{\sigma=\pm} \sigma {2\kappa \dot\phi_0^2 \over 3}
      \delta(y-y_{\sigma})  K^{(J)}_{\sigma}
    +  \epsilon ^{(J)} S_Y  \,,
\end{eqnarray}
with
\begin{eqnarray}
    K^{(J)}_{\sigma}(r)
 :=   2 a_{\sigma}^2  
    \left[
    \stac{(J)}{ \hat\xi^y_{\sigma}}
 - \sigma \frac{3 \lambda_{\sigma} \Delta Y^{(J)} }
          {2 \kappa a_{\sigma}^2 \dot\phi_0^2}    
+ \frac{\epsilon^{(J)} }{\dot\phi_0 }
  \left(
        \frac{\lambda_{\sigma}}{2} S_{jun}^{\sigma}
    - \frac{3}{2\kappa \dot \phi_0} 
    \Bigl[ S_\varphi + \partial_y  \Delta^{-1} 
      S_{\psi}  
    \Bigr]
  \right)
   \right]_{y=y_{\sigma}}  \,.
 \nonumber
\end{eqnarray}
Solving these equations for $A^{(J)}$ and $Y^{(J)}$ as well as
Eq.~(\ref{eq: Lap xi^y = EM}) for $\hat\xi^y_\pm$ , and using the gauge transformation Eq.~(\ref{eq:gauge trans ABC}) with the aid of Eq.~(\ref{eq:TT cond}), we obtain $\bar A^{(J)}$, $\bar B^{(J)}$, and $\bar C^{(J)}$, which represent the metric perturbations induced on the branes.

    \section{Gradient Expansion}
    \label{sec:mode}

    \subsection{Green's function}

We can write down the formal solution of Eq.~(\ref{eq:A^J with Junc}) by means of the Green's function, 
\begin{eqnarray}
     a^2 A^{(J)} =2 \kappa \sum_{\sigma=\pm} \int d^3x ~ G_A 
     \Sigma^{(J )}_{\sigma} 
     + 2 \epsilon^{(J)} 
     \int d^3x \int^{y_{-}}_{y_{+}} dy G_A S_A \,,
\label{eq:solv A^J}
\end{eqnarray}
where the $G_A({\mathbf x},y;{\mathbf x'},y')$ is the Green's function for the TT part in the static case. The factor 2 in the second term reflects the $Z_2$ symmetry of this brane world model.
In the static case, the Green's function is given by
\begin{eqnarray}
    G_A({\mathbf{x}},y;{\mathbf{x}}',y') 
&=& - 
\int  \frac{d^3 {\mathbf k}}{(2\pi)^3}  
e^{i{\mathbf{k}}  ({\mathbf x}-{\mathbf x'})}
    \left[
        \frac{N a(y)^2 a(y')^2 }{ {\mathbf k}^2+\epsilon^2}
     +  \sum_i  \frac{w_i(y) w_i(y')}{m_{Ki} ^2+ {\mathbf k}^2 }  
   \right] \,,
\label{eq:static Green fun. G} 
\end{eqnarray} 
where $w_i(y)$ is the mode function, and its orthonormal conditions are given by
\begin{eqnarray}
    \int _{y_{+}} ^{y_{-}} w_i dy =  0 \,,
\qquad
 2 \int _{y_{+}}^{y_{-}} 
     \frac{ w_i w_j  }{a^{2}}  dy = \delta_{ij} \,.
\label{eq:orthogonality}
\end{eqnarray}
The normalization factor $N$ is defined by 
\begin{equation}
    N:= \left[2\int_{y_{+}}^{y_{-}} a^2 dy\right]^{-1} \,.
\label{eq:defN}
\end{equation}
The first term on the right hand side of Eq.~(\ref{eq:static Green fun. G}) is the contribution from the zero mode whose four-dimensional mass eigenvalue is zero. The second term corresponds to the propagator due to the Kaluza-Klein (KK) excitations whose \textit{i}th excitation has the discrete mass eigenvalue $m_{K i}$. We refer to these modes as KK modes.

As for the scalar-type perturbations, it has been proved in linear perturbation by considering the source free equation that there is no physical mode with a zero eigenvalue of the four-dimensional D'Alembertian \cite{Tanaka:2000er}. This means that the massless scalar-type mode disappears when the stabilization mechanism is taken into account. 
The explicit mode function for the lightest mass mode is found in Ref.\cite{Tanaka:2000er}.

    \subsection{Transverse-traceless perturbation} 

    \subsubsection{Temporal component} 

By the zero mode truncation, in which we substitute only the first term in Eq.~(\ref{eq:static Green fun. G}) into Eq.~(\ref{eq:solv A^J}),  we obtain  
\begin{eqnarray}
   \Delta A^{(J)}_{0} (r,y)
   = - 2 \kappa N \sum_{\sigma=\pm} a_{\sigma}^4
    (T^0_0-\frac{1}{3}T)^{(J)}_{\sigma}  
    + 2N \epsilon^{(J)} 
    \left[
        \int^{y_{-}}_{y_{+}}  a^2 S_{A}  dy
    -   \sum_{\sigma=\pm} \sigma a_{\sigma}^4
        S_{\Sigma}^{\sigma} 
    \right] \,,
\label{eq:A^J_0}
\end{eqnarray}
where we did not assume any truncation for the source terms $S^\pm_\Sigma$ and $S_A$. We assigned the label $0$ to indicate the zero mode truncation.

To evaluate the contribution from the second term in Eq.~(\ref{eq:static Green fun. G}) for the TT part of the metric perturbations, we follow the strategy that is used in Ref.\cite{Tanaka:2000er}. 
Rewrite the part coming from KK modes in the Green's function as 
\begin{eqnarray}
   \int  \frac{d^3 {\mathbf k}}{(2\pi)^3}  
   e^{i{\mathbf{k}}  ({\mathbf x}-{\mathbf x'})}
   \sum_i \frac{w_i w_i}{m_{Ki} ^2+ {\mathbf k}^2 }  
 &=&
    \sum_i \frac{w_i w_i }{m_{Ki}^2}
        \delta^3({\mathbf{x}-\mathbf{x}'})
  - \sum_i \frac{w_i w_i }{m_{Ki}^2}
    \int \frac{dk^3}{(2\pi)^3}
      \frac{ {\mathbf k}^2  e^{i {\mathbf{k (x-x')}}} }{m_{Ki}^2+ {\mathbf k}^2}  \,.
\end{eqnarray}
Under the condition that $k^2/m_{Ki}^2 \ll 1$ holds, the first term on the right hand side gives the dominant contribution. 
Notice that the first term is nothing but the Green's function for Eq.~(\ref{eq:Ein-A^J}) with $\Delta=0$. Thus to pick up this part of the Green's function is equivalent to solving the equation for $A^{(J)}$ by setting $\Delta=0$ from the beginning. Substituting $A^{(J)} \equiv A^{(J)}_0+A^{(J)}_S$, where $A_0$ is the zero mode part and $A_S$ is the KK mode part, into 
Eq.~(\ref{eq:A^J with Junc}) and neglecting the $\Delta$ term for the  KK mode contribution, we obtain
\begin{eqnarray}
    \hat L^{(TT)} \bigl( a^2 A^{(J)}_{S} \bigr)
    &\approx& 
    - 2 \kappa   
    \sum_{\sigma=\pm} \Sigma^{(J)}_{\sigma}
    \biggl[
        N a_{\sigma}^2 - \delta(y-y_{\sigma})
    \biggr]
 + \epsilon^{(J)}  {\mathbb S}_A \,,
\end{eqnarray}
where 
\begin{eqnarray}
   {\mathbb S}_A  := 
    S_{A} -2N \int^{y_{-}}_{y_{+}} a^2 S_{A } dy \,.
\end{eqnarray}
Applying the integration operator $\int^{y} a^{-4}\int^{y}_{-y}dy
a^2$, this equation is formally solved as 
\begin{eqnarray}
    A^{(J)}_{S} (r,y)
   = - 2 \kappa N   
    \sum_{\sigma=\pm} a_{\sigma}^2 \Sigma^{(J)}_{\sigma}
    \biggl(
      \int^{y}_{y_{-\sigma}} dy' \frac{ v_\sigma }{a^2 }
     -C_{\sigma} 
    \biggr)
 + \epsilon^{(J)} \Bigl[  
        \int^{y}_{y_{+}} \frac{dy'}{a^4 }
       \int^{y'}_{y_{+}} dy'' a^2 \, {\mathbb S}_A - D(r) 
   \Bigr]\,,
\label{eq:KK mode A^J}
\end{eqnarray}
where $C_+$ and $C_-$ are constants and $D (r)$ is a function of $r$, which are determined later. The function $v_\pm(y)$ is defined by 
\begin{equation}
    v_{\pm}(y):={1\over a ^2}\int_{y_{\mp}}^y dy' a(y')^2 \,.
    \label{eq:def. v}
\end{equation}
From the orthogonality of Eq.~(\ref{eq:orthogonality}), 
$A^{(J)}_{S}$ must be orthogonal to the zero mode function.
This condition fixes the constants $C_{\pm}$ and the function $D(r)$ as 
\begin{eqnarray}
  C_{\sigma} &=&  2N
  \int^{y_{-}}_{y_{+}} dy \, a^2 
  \int^{y}_{y_{-\sigma}}  {dy'} 
  \frac{v_{\sigma} }{a^{2} } \,,
\nonumber
\\
   D (r) &=& 2N
    \int^{y_{-}}_{y_{+}} dy \, a^2 
    \int^{y}_{y_{+}}  \frac{dy'}{ a^{ 4} }
      \int^{y'}_{y_{+}} dy'' a^2 \,  {\mathbb S}_A  \,.
\end{eqnarray}
Integrating by parts and taking $y=y_{\pm}$, Eq. (\ref{eq:KK mode A^J}) is reduced to 
\begin{eqnarray}
    A^{(J)}_{S} (r,y_{\pm})
   &=& - 4\kappa N^2    
       \sum_{\sigma=\pm} a_{\sigma}^2 \Sigma^{(J)}_{\sigma}
      \int^{y_{-}}_{ y_{+} } dy' 
        v_{\sigma}^2
   \pm 2\kappa N a_{\mp}^2 \Sigma^{(J)}_{ \mp } 
   \int^{y_{-}}_{y_{+}} \frac{v_{\mp}}{a^2} dy
\nonumber
\\ 
&& + 
     2N \epsilon^{(J)}
     \int^{y_{-}}_{y_{+}} dy' \frac{v_{\pm}}{a^2 }
     \int^{y'}_{y_{+}} dy'' a^2 {\mathbb S}_A \,.
\label{eq:A_[GGK] at y_+-}
\end{eqnarray}
where we used Eqs.~(\ref{eq:dot u v}) and (\ref{eq:u v on y+-}).

The approximation method that we have used is a kind of derivative expansion method, in which the typical wavelength of perturbations is supposed to be long. This expansion is valid only when the smallest mass of the KK excitations is sufficiently large. 
We can obtain the higher order corrections by iteration, in which the $\Delta$ term, which we have neglected in the above discussion, is incorporated as the source term. For the TT part, we do not consider further iteration than Eq.~(\ref{eq:KK mode A^J}). As we will see later, in the scalar-type perturbations, we need to consider one iteration of the 
$\Delta$ term to obtain results accurate to the same order.

 \subsubsection{ Spatial component }

Let us turn to the spatial components of the TT part. 
Since each spatial component depends on the gauge choice of $\hat\xi^r$, it is convenient to deal with the gauge invariant combination \cite{Kudoh:2001wb} 
\begin{eqnarray}
  \Bigl[ \partial_r (r \bar C^{(J)}) - \bar B^{(J)} \Bigr]_{y=y_\pm}
&=& 
- r \partial_r 
   \Bigl[ 
     \frac{1}{2}  \bar A^{(J)} 
    + \frac{3}{2}  (\psi^{(J)} + 2H \stac{(J)}{\xi^y })
    + \epsilon^{(J)} S_B
   \Bigr] \,,
      \qquad (y=y_\pm)
   \,,
\label{eq:gauge inv comb}
\end{eqnarray}
where we used Eqs.~(\ref{eq:TT cond}) and (\ref{eq:gauge trans ABC}), and  $S_B$ is given by
\begin{eqnarray}
 S_B 
 &=& - \frac{\hat \xi^y}{2}  
        \bigl( B_{,y}+2\kappa \dot\phi_0 \varphi \bigr)
   - \frac{\hat\xi^r}{2}  \bigl[ B+3(Y+2H \hat \xi^y)\bigr]_{,r}
   - \frac{3}{2} \dot H (\hat\xi^y )^2 
   + \frac{(\hat\xi^r )^2}{r^2} 
   + \frac{1}{r} \partial_r 
    \Bigl[ \frac{ (\hat\xi^y )^2}{a^2}- (\hat\xi^r )^2 \Bigr]
\nonumber \\
 & & +
 \int \frac{dr}{r}
  \biggl\{  \hat\xi^r (\bar C- \bar B)_{,r} 
   + \hat\xi^y
  \Bigl[ 
    \frac{r B_{,ry}}{2} - \frac{2}{a^2} \Bigl( \hat\xi^y_{,rr}- \frac{1}{r} \hat\xi^y_{,r}\Bigr)
  \Bigr] 
  + (\hat\xi^r_{,r})^2 
  - \frac{ (\hat\xi^r)^2}{r^2}
  - \frac{ (\hat\xi^y_{,r})^2 }{a^2} 
  \biggr\} \,.
\end{eqnarray}
To fix the gauge degrees of freedom corresponding to the choice of the radial coordinate, we adopt the isotropic gauge that is defined by $\bar B=\bar C$ on each brane. With this choice of gauge, the left hand side of Eq.~(\ref{eq:gauge inv comb}) becomes $r \partial_r \bar B$, and  we immediately obtain the spatial component $\bar B$ $ (= \bar C)$ in Gaussian normal coordinates: 
\begin{eqnarray} 
   \Delta \bar B^{(J)}  (r,y_\pm )
   &=& - \frac{1}{2} \Delta \bar A^{(J)}  
 \mp \frac{\kappa H}{2}a^2_\pm T^{(J)}_\pm
 - \frac{3}{2} \Delta Y^{(J)}
 - \frac{3}{2} \epsilon^{(J)}
\left(
  \frac{2}{3}\Delta  S_B 
  +S_\psi + 2Ha^2_\pm S^\pm_\xi
\right)\,.
\label{eq:bar B^2 rew 1}
\end{eqnarray}
It is also necessary to specify the explicit form of the radial gauge parameter at first order, 
$\stackrel{\scriptscriptstyle(1)\ }{\hat\xi_{\pm}^r}$, since the second order perturbations $\bar A^{(2)}$ and $\bar B^{(2)}$ depend on it. 
Substituting Eq.~(\ref{eq:gauge trans ABC}) into the isotropic gauge condition, $\stackrel{\scriptscriptstyle(1)\ }{\hat\xi_{\pm}^r}$ is determined as
\begin{eqnarray} 
    \stackrel{\scriptscriptstyle(1)\ }{\hat\xi_{\pm}^r} (r)
     = -\frac{r}{4} { B }^{(1)}(r, y_{\pm}) \,.
\label{eq:iso gauge} 
\end{eqnarray}

    \subsection{Scalar-type perturbation}

As mentioned earlier, there is no zero mode in the scalar-type perturbation. To evaluate the contribution from massive modes, we apply the same technique that we used in the preceding subsection for the KK modes of the TT part.
First we consider the equation for $Y^{(J)}$ by setting $\Delta=0$.  The homogeneous solutions of 
Eq.~(\ref{eq:Ein-Y^J}) with  $\Delta=0$ are given by 
\begin{equation}
 u_{\pm}(y)=1-2 H(y) v_{\pm}(y), 
\label{eq:def. u_+-}
\end{equation}
where $v_{\pm}(y)$ is defined in Eq.~(\ref{eq:def. v}). 
The Green's function $G_Y$, which satisfies 
\begin{eqnarray}
 \frac{1}{a^2\dot\phi_0 ^2} \hat L^{(Y)}
 \Bigl( \frac{G_Y}{a^2}\Bigr)
 =- \delta (y-y') \,,
\end{eqnarray}
is constructed as 
\begin{equation}
  G_Y(y;y') = \frac{3Na^2(y) a^2(y')}{\kappa} 
  \Bigl[
    u_-(y) u_+ (y') \theta(y'-y) + u_-(y') u_+ (y) \theta(y-y')
  \Bigr]\,.
\end{equation}
Then using this Green's function, 
Eq.~(\ref{eq:Y^J with Junc}) with $\Delta=0$ is solved as 
\begin{eqnarray}
    Y^{(J)} (r,y) &=& 
   - N \sum_{\sigma={\pm}} \sigma u_{\sigma}(y)
\left[
     K_{\sigma}^{(J)} 
    +
     \epsilon^{(J)}  
     \int_{y_{\sigma}} ^y  
    \frac{3 u_{-\sigma} S_Y}{\kappa \dot\phi_0^2} dy' 
\right] \,.
\label{eq:Y^2 -  approx}
\end{eqnarray}
We assign the label $0$ because this term gives a contribution to the metric perturbation at the same order as the zero mode in the TT part of the perturbation, although it is related to massive scalar-type modes.

Using Eqs.~(\ref{eq:sum v u etc}) and (\ref{eq:formula of S_psi}), it becomes 
\begin{eqnarray}
 \Delta  Y^{(J)}_{0}(r, y) 
    &= &
  - \frac{\kappa N }{3} \sum_{\sigma=\pm} a_{\sigma}^4 
    T^{(J)}_{\sigma}  u_\sigma(y)
  - \epsilon^{(J)} S_\psi 
- 2N  \sum_\sigma  \sigma  \frac{u_\sigma (y)}{H(y_\sigma)}
  \Delta L_\sigma^{(J)}
\nonumber
\\
 && 
  - 2N \epsilon^{(J)} \sum_\sigma \sigma u_\sigma (y)
\biggl[
     a_{\sigma}^4 S_{\xi}^{\sigma}
 +  \int^{y}_{y_{\sigma}} dy 
   \Bigl(
    a^2 v_{-\sigma } \Delta  {\mathbb S}_\varphi
 +   \frac{3u_{-\sigma} }{ 2\kappa\dot \phi_0^2} 
    \Delta  {\mathbb S}_Y  
 -  a^2 S_\psi
    \Bigr) 
\biggr]   \,,
\label{eq:zero mode of Y^J(r,y)}
\end{eqnarray}
where 
\begin{eqnarray}
L_\sigma^{(J)}(r) := H(y_\sigma)
  \biggl[ 
    - \sigma \frac{3 \lambda_{\sigma}}{2\kappa\dot\phi_0^2} \Delta Y^{(J)}
   + \epsilon^{(J)} a_{\sigma}^2
  \Bigl(
    \frac{ \lambda_{\sigma}}{2\dot\phi_0} S^{\sigma}_{jun} 
  - \frac{ \ddot\phi_0  \varphi^2}{2 \dot\phi_0^3} 
  + \frac{3 \varphi  \Delta  Y }
      {2\kappa a^2 \dot\phi_0^3}
  \Bigr)
   \biggr] _{y=y_{\sigma}}\,.
\label{eq:L_sigma}
\end{eqnarray}
As we can see from the first term of Eq.~(\ref{eq:zero mode of Y^J(r,y)}), this mode of the scalar-type perturbation partly gives long-ranged contributions, as we anticipated. We refer to this part of the scalar-type perturbation as the pseudo-long-ranged part to distinguish the remaining short-ranged correction. 
We use the subscript $S$ to represent the short-ranged part, although it is also used for the KK mode.

The source term for the next order correction $Y _{S}$ is given by $\Delta Y_{0}$, which we neglected in the calculation of the pseudo-long-ranged part.
Since the Green's function $G_Y$ is already known, we easily obtain 
\begin{eqnarray}
  Y^{(J)}_{S}(r,y)&=& 
  \frac{3N}{\kappa} 
  \sum_{\sigma={\pm}} \sigma u_{\sigma}(y)
     \int_{y_{\sigma}} ^y  
     \frac{u_{-\sigma} \Delta Y ^{(J)}_{0}}{\dot\phi_0^2}  dy' \,.
\label{eq:Y^J - correction}
\end{eqnarray}
Setting $y=y_{\pm}$ and using Eqs.~(\ref{eq:u v on y+-}) and (\ref{eq:rel-Sum u}), the expressions for $Y_{0}$ and $Y_{S}$ are summarized as 
\begin{eqnarray}
 \Delta Y^{(J)}_{0} (r, y_{\pm})
   &=& 
  - \frac{\kappa N}{3} \sum_{\sigma=\pm} a_{\sigma}^4 
    T^{(J)}_{\sigma}  
    \mp \frac{\kappa H}{3} a_{\pm}^2 T_{\pm}^{(J)}
 -\epsilon ^{(J)} 
 \Bigl(  S_\psi + 2H a_{\pm}^2 S_{\xi}^{\pm}
 \Bigr) 
    - 2N \epsilon^{(J)} \biggl[
     \sum_{\sigma=\pm} \sigma a_{\sigma}^4
     S_{\xi}^{\sigma} 
\nonumber
\\
&&
   + \int^{y_{-}}_{y_{+}} dy 
     \Bigl(
  a^2 v_{\pm} \Delta {\mathbb S}_\varphi 
    + \frac{3u_{\pm}} {2\kappa \dot \phi_0^2} \Delta {\mathbb S}_Y 
    -a^2 S_\psi 
  \Bigr)
\biggr]
 - 2N  \sum_\sigma \sigma  \frac{u_\sigma (y_\pm)}{H(y_\sigma)}
  \Delta L_\sigma^{(J)} \,,
\label{eq:Y^J_[GG0] y_+-}
\\
    Y^{(J)}_{ S} (r, y_{\pm})
 &=&  \frac{3N}{\kappa}
       \int_{y_{+}}^{y_{-}}  
      \frac{u_{ \pm } \Delta Y ^{(J)}_{ 0} }
      { \dot\phi_0^2}   dy'  \,. 
 \label{eq:Y^J_[GGS] y_+-}
\end{eqnarray}

 \subsection{ Large coupling limit }
\label{sec:lcl}
In the preceding sections, we derived the formal solutions to evaluate the second order perturbations. However, the result is very complicated. To simplify the analysis, we assume $|V_{(\pm)}''| \gg |\ddot\phi_0 / \dot\phi_0|$, and take the limit 
\begin{eqnarray}
    \lambda_{\pm} \to 0 .
\label{eq:large limit}
\end{eqnarray}
In the case of the Goldberger-Wise stabilization model \cite{Goldberger:1999uk}, this limit corresponds to their large coupling constant.

In this limit, the junction condition (\ref{eq:Jun of varphi}) for $J=1$ becomes
\begin{eqnarray}
    \varphi^{(1)} (r,y_{\pm}) \approx 
    \dot\phi_0 \stac{(1) }{\hat\xi_{\pm}^y } \,.
\label{eq:varphi^1 at y_pm}
\end{eqnarray}
Here we mention that the source term $S^{\pm}_{jun}$ which is given by Eq.~(\ref{eq:Jun of varphi}) contains $V''_{(\pm)}$ and $V'''_{(\pm)}$. Hence $\lambda_\pm S^\pm_{jun}$ does not vanish even in this limit, and it is reduced to
\begin{eqnarray}
    \frac{ \lambda_{\pm}}{2} S^{\pm}_{jun} 
  \approx
    ( \delta \varphi^{(2)} - \dot\phi_0 \stac{(2)}{\xi^y} )\,,
\label{eq:S_jun approx}
\end{eqnarray} 
where we used Eq.~(\ref{eq:varphi^1 at y_pm}). Therefore the junction condition for the scalar field in this limit is summarized as 
\begin{eqnarray}
    \varphi^{(J)} -  \delta \varphi^{(J)} 
    \approx 0 \quad  (y=y_{\pm}) \,.
\label{eq:varphi^J at y_pm}
\end{eqnarray}
Under this condition, the last terms in Eqs.~(\ref{eq:S_Sigma}) and (\ref{eq:S_xi}) vanishes.
In particular, Eq.~(\ref{eq:S_jun approx}) gives 
\begin{eqnarray*}
   \frac{ \lambda_{\pm}}{2\dot\phi_0}
   S^{\pm}_{jun} 
=
 \left( 
    \frac{ \ddot\phi_0   \varphi^2}{2 \dot\phi_0^3} 
  - \frac{3 \varphi \Delta  Y }
      {2\kappa a^2 \dot\phi_0^3}
\right)_{y=y_{\pm}}\,,
\end{eqnarray*}
and then we obtain the approximation for (\ref{eq:L_sigma}) 
\begin{eqnarray}
    L^{(J)}_{ \pm} \approx 0 \,.
\label{eq:L=0}
\end{eqnarray}

    \section{Recovery of the 4D Einstein Gravity: First order}
    \label{sec:4D recovery-1st}

    \subsection{Linear perturbation}

We review the results for linear perturbations in terms of the notation of the present paper. From Eq.~(\ref{eq:A^J_0}), the zero mode truncation of the TT part is given by
\begin{eqnarray}
    A^{(1)}_0
 =
    \frac{8}{3} \sum_{\sigma=\pm} \Phi_\sigma \,,
 \quad
   B^{(1)}_0= - \frac{8}{3r}\partial_r \Delta^{-1} \sum_{\sigma=\pm} \Phi_\sigma \,,
\label{eq:A^1 zero mode} 
\end{eqnarray}
where we have introduced the Newton potential by 
\begin{eqnarray}
    \Delta \Phi_\pm (r)
    &:=&
      4 \pi G ~ \rho^{(1)}_{\pm}(r)   \,,
\label{eq:def Newton potential}
\end{eqnarray}
and $G$ is the induced four-dimensional Newton's constant defined by 
\begin{eqnarray}
      8\pi G &:=& \kappa N  \,. 
\label{eq:4D Newton cons}
\end{eqnarray}

{}From Eq.~(\ref{eq:zero mode of Y^J(r,y)}), the pseudo-long-ranged part in the scalar-type perturbation is obtained as 
\begin{eqnarray}
   Y^{(1)}_0 = \frac{2}{3} \sum_{\sigma=\pm} u_{\sigma}(y) \Phi_\sigma(r) \,,
   \quad
   \varphi^{(1)}_0  = \frac{2}{3} \dot \phi_0 \sum_{\sigma=\pm} v_{\sigma}(y) \Phi_\sigma(r) \,,
\label{eq:Y^1 zero mode} 
\end{eqnarray}
and the gauge transformation is 
\begin{eqnarray}
    {\hat \xi^y_\pm }= \mp \frac{\Phi_\pm}{3N a_{\pm}^2} \,.
    \label{eq:xi^y - Phi}
\end{eqnarray}

In this paper we concentrate on the gravity on one of the $Z_2$ symmetric branes that carries matter fields on it as a source of gravitational field. We assume that the energy-momentum tensor of the matter fields on the other brane vanishes. By this simplification, the sum of $\Phi_+$ and $\Phi_-$ is replaced by
\begin{eqnarray}
   \sum_\sigma \Phi_\sigma 
   \to 
   \Phi_\pm \,,
\label{eq:simp} 
\end{eqnarray}
in each case. 
Here we note that, to avoid confusion in showing the formulas for two different situations simultaneously, we are using a different convention for the physical length scale from that used in Ref.~\cite{Tanaka:2000er}.

Substituting the above formulas into Eq.~(\ref{eq:gauge trans ABC}), we obtain 
\begin{eqnarray}
\bar A^{(1)}_{0\pm} (r,y_\pm)
  &=&  2 \Phi_\pm \,, 
\end{eqnarray}
where we attached a subscript $\pm$ on the perturbation quantities to specify in which case we are working. For instance, $\bar A_{+}$ represents the value of $\bar A$ when only the matter fields on the positive tension brane are taken into account. The remaining metric functions turn out to be 
\begin{eqnarray}
    \bar B_{0\pm}^{(1)}= \bar C_{0\pm}^{(1)} = -2 \Phi_\pm 
\end{eqnarray}
in the isotropic gauge (\ref{eq:iso gauge}).
These results coincide with the results for the four-dimensional Einstein gravity.

 \subsection{Correction to the leading term}

To obtain an approximate estimate for the corrections due to the KK mode or the short-ranged part of the scalar-type perturbations, it is useful to consider cases in which the back reaction of the bulk scalar field on the background geometry is weak; namely,  
\begin{eqnarray}
    \frac{|\dot H|}{H^2} = \frac{ \kappa \dot\phi_0^2}{3H^2} \,, 
\nonumber
\end{eqnarray}
is not as large as unity. 
In a weak back reaction, the metric is approximately given by the pure anti-de Sitter form
\begin{eqnarray}
    a (y) \approx e^{-y/\ell} \,,
\end{eqnarray}
and we set 
\begin{eqnarray}
    y_{+}=0, \quad y_{-} = d \,.
\end{eqnarray} 
Here $\ell$ is the curvature radius of AdS$_5$.

If we substitute this warp factor, 
Eqs. (\ref{eq:def. v}) and  (\ref{eq:def. u_+-}) are approximated as  
\begin{equation}
 u_{\pm}(y) \approx \frac{a_{\mp}^2}{a^2} \,,
\quad
 v_{\pm}(y) \approx  \frac{\ell}{2} 
 \Bigl( \frac{a_{\mp}^2}{a^2} - 1 \Bigr) \,.
\label{eq:u v in W.B.}
\end{equation}
Here one remark is in order. 
The above expression for $u_+$ is not a good approximation near the positive tension brane because $u_+(y_+)$ depends on the difference between $N$ defined in Eq. (\ref{eq:defN}) and $H$ at $y=y_+$. 
The value of $N$ in the weak back reaction limit is 
$H(a_+^2-a_-^2)^{-1}$, and the difference between $H$ and $N$ is hierarchically suppressed. However, unless we consider an extreme case, the deviation of the value of $N$ from this limiting value is not hierarchically small. As a result, we have  
\begin{equation}
 u_{+}(y_+) = \sO (1) \,, 
 \label{eq:u_+(y_+)}
\end{equation}
instead of $\sO (a_-^2)$. 
This also means that the $y$ dependences of $Y_{0+}^{(1)}$ and $Y_{0-}^{(1)}$ are different. If the single mode with the lowest mass eigenvalue dominates the scalar-type perturbation, the $y$ dependence must be the same for both cases. Therefore, we find that the modes with higher mass eigenvalues also contribute to the behavior of $Y_{0+}^{(1)}$ near the positive tension brane. 

Let us consider the KK mode contribution (\ref{eq:KK mode A^J}) in the linear perturbation. A straightforward calculation shows\cite{Tanaka:2000er}
\begin{eqnarray}
   A^{(1)}_{S\pm} (r,y)
    \approx  - 
   \frac{\ell^2  \Delta  \Phi_{\pm}}{3}
   \biggl[ \frac{a_{\mp}^2}{a^4}
         - \frac{2}{a^2} +
        \biggl(\frac{4 d}{ \ell } -\frac{1}{a_\pm}
        \biggr) 
   \biggr] \,,
\label{eq:A^1_S(y,r)}
\end{eqnarray}
where we have assumed $d/\ell \gg 1$.
Hence, on the brane where the matter fields reside, it becomes 
\begin{eqnarray}
 &&   A^{(1)}_{S+}(r,y_+) 
   \approx 
   { \ell^2 (3-4d/\ell)  \over 3 a_{+}^4} 
    { \Delta  \Phi_{+}} 
   \,,   
\nonumber
\\ 
 &&  A^{(1)}_{S-} (r,y_-)
    \approx 
  - {\ell^2 \over 3 a_{-}^4 } {\Delta \Phi_{-} }      
\,.
\label{eq:KK correction}
\end{eqnarray}
To compare the KK mode contribution with the zero mode one, we evaluate the ratio between them. Then we find that the KK mode contribution is suppressed by the factor 
\begin{eqnarray}
 \beta_\pm
   :=   \frac{\ell^2}{a^4_\pm  r_\star^2} 
    = \left\{
\begin{minipage}[c]{8cm}
$\displaystyle
   \hspace{50pt} \frac{\ell^2}{r_\star^2},  
   \hspace{50pt} (y=y_+) \,,$
\newline
\newline
$\displaystyle
    \Bigl( \frac{ 0.1 {\mathrm{mm}}}{r_\star} \Bigr)^2
    \Bigl( \frac{ 10^{-16} }{a_-}  \Bigr)^4
    \Bigl( \frac{ \ell}{\ell_{Pl}}  \Bigr)^2,
     \quad ( y=y_-),$
\end{minipage} 
\right.
\label{eq:suppres factor 1}
\end{eqnarray}
where we introduced a typical length scale $r_\star$, and performed 
a replacement like $\Delta \approx r_\star^{-2}$.

On the positive tension brane the KK mode contribution 
is suppressed at $r_\star \gg \ell$ at linear order. 
Note that, if one takes the limit $d/\ell \to \infty$, the KK mode 
(\ref{eq:KK correction}) seems to diverge. 
In this limit the lowest KK mode mass goes to zero, and the mass spectrum becomes continuous. Then, our expansion scheme which we call a gradient expansion, is no longer a good approximation. 
Hence, this divergence in the large $d$ limit is just due to the breakdown of our expansion scheme.

On the negative tension brane the KK mode becomes dominant only at the length scale $\lesssim 0.1$ mm when the AdS curvature length $\ell$ and the hierarchy $a_+/a_-$ are set to the Planck length $\ell_{Pl}$ and $10^{16}$, respectively. 
One may think that the deviation from four-dimensional Einstein gravity at the sub-millimeter scale provides an observable effect. 
However, in the KK mode contribution to the gravitational potential, 
$\Phi_{\pm}$ appears only in the form of 
$(4\pi G)^{-1}\Delta\Phi_{\pm}$, which is equal to the matter energy density $\rho_{\pm}$. Hence, the KK mode does not contribute to the force outside the matter distribution. 
Therefore, this effect appears to be hard to observe.

Next, we consider the short-ranged part of the scalar-type perturbation.  
The short-ranged part (\ref{eq:Y^J - correction}) in the weak back reaction limit is evaluated as
\begin{eqnarray}
   Y^{(1)}_S (r,y)
   \approx 3N a^2_- a^2_+ \Delta Y_0^{(1)}
           \int^{y_-}_{y_+} \frac{dy}{a^4 \kappa \dot\phi_0^2}
   \approx {1\over a_-^2 \tilde  m_S^2}\Delta Y_0(r,y), 
\label{eq:Y_S-1}
\end{eqnarray}
where  $a^2 Y_0^{(1)}\approx \mathrm{const}$ is used (Appendix \ref{sec:Integration, GW}). 
Here we introduced the lowest mass eigenvalue in the scalar-type perturbation $m_S:=a_-\tilde m_S$, whose order of magnitude is determined by the last equality\cite{Tanaka:2000er}. 
We refer to $m_S$ as the radion mass. The reason why $m_S$ defined above gives the lowest mass eigenvalue can be understood as follows. 
Suppose the mode with the lowest mass squared dominates perturbations in the long-wavelength limit. Then the propagator for the scalar perturbation should be proportional to $1/(\Delta-m_S^2)$. In our approximation of a gradient expansion, this massive propagator is expanded as $(1/m_S^2)+(\Delta/m_S^4)+\cdots$. The first term gives 
$Y^{(1)}_0$ and the second $Y^{(1)}_S$. Hence, the ratio between them is $\Delta/m_S^2$. However, again this simple-minded estimate is not correct for $Y_{S+}^{(1)}$ near the positive tension brane for the same reason that the approximate expression for $u_+$ [Eq. (\ref{eq:u v in W.B.})] is not valid near the positive tension brane. 
In fact, the value of $Y_{S+}^{(1)}$ on the brane can be evaluated by using Eq.~(\ref{eq:Y^J_[GGS] y_+-}). Substituting the estimate given in Eqs.~(\ref{eq:approx int}), we obtain 
$ Y^{(1)}_{S+} (r,y_+)\approx {\Delta Y^{(1)}_{0+}/\tilde m_S^2}$.
Taking this into account, we guess that the formula (\ref{eq:Y_S-1}) should be modified as 
\begin{eqnarray}
   Y^{(1)}_{S\pm} (r,y)
    \approx \frac{1}{ \tilde m_S^2 a_{\pm}^2 a^2} \Delta
    \Phi_{\pm}(r) \,.   
\label{Eq:YScorrect}
\end{eqnarray}
We give a justification of this formula in Appendix \ref{sec:Integration, GW}.
Then, the ratio $Y^{(1)}_{S\pm} (r,y_\pm)/A^{(1)}_{0\pm} (r,y_\pm)$, where we compare the short-ranged part to the Newtonian potential, is 
\begin{eqnarray}
\gamma_\pm :=
     \frac{1}{ a_{\pm}^4 \tilde m_S^2 r_\star^2}
     = \beta_\pm \Bigl( \frac{1}{\tilde m_S \ell} \Bigr)^2 \,.
\label{eq:order m_S r}
\end{eqnarray}
When the radius stabilization mechanism proposed by Goldberger and Wise works most efficiently, the mass $\tilde m_S$ becomes $O( \ell^{-1})$\cite{Tanaka:2000er,Goldberger:1999uk}.  
In this case the short-ranged part of the scalar-type perturbation is suppressed for the same reason as the KK mode. 
We have shown than the zero mode and pseudo-long-ranged part reproduce the correct four-dimensional Einstein gravity. The remaining KK mode and the short-ranged part accompany extra suppression factors 
$\beta_{\pm}$ and $\gamma_{\pm}$, respectively.

    \section{Recovery of the 4D Einstein Gravity: Second order}
    \label{sec:4D recovery-2nd}

We discuss the second order perturbations in the large coupling limit discussed in Sec. \ref{sec:lcl}.  
As in the case of the first order perturbation, we iteratively solve the equations of motion by using gradient expansion. In the equations we derive below, we neglect the terms that are relatively suppressed by the factor of 
$1/r_\star^4$ compared to the leading contribution. For convenience, we quote the contribution from the zero-type coupling, which is obtained by substituting Eqs.~(\ref{eq: phi^J=Y^J+...}), 
(\ref{eq:xi^y}), (\ref{eq: Lap xi^y = EM}), (\ref{eq:A^J_0}), and (\ref{eq:Y^J_[GG0] y_+-}) into Eq.~(\ref{eq:gauge trans ABC}):
\begin{eqnarray}
  {\Delta}  \bar A^{(2)} _{0\pm}(r,y_{\pm})
   &=&  8 \pi G 
      (\rho ^{(2)}_{\pm} + 3P ^{(2)}_{\pm}  )
 - {\Delta} 
   \Bigl[
     {\hat\xi_{ \pm }^y} \bar{A} _{\pm,y} 
   + {\hat\xi_{ \pm }^r} \bar{A} _{\pm,r}
   +  
\underline{ \dot H ( {\hat\xi_{ \pm }^y})^2}
    \Bigr] 
 \pm 2N  a_{\pm}^4 (S_{\xi}^{\pm} - S_{\Sigma}^{\pm})
\nonumber
\\
& &  
 +  {2N}  \int^{y_{-}}_{y_{+}} dy 
  \biggl[
    a^2 ( S_{A\pm} - S_{\psi\pm})
  + a^2 v_{\pm} \Delta {\mathbb S}_{\varphi\pm} 
  + \frac{3u_{\pm}}{2 \kappa \dot \phi_0^2}  
    \Delta {\mathbb S}_{Y\pm}
\biggr] \,.
\label{eq:A_[GG0] rew}
\end{eqnarray}
Here we have used the fact that $S_\xi$ and $S_\Sigma$ on the brane without matter distribution vanish, which is easily verified just by noticing that $\stac{(1)}{\hat\xi_{\mp}^y} = 0$ and $\bar A_{\pm,y}^{(1)} (y_{\mp}) = 0.$  
The contribution from $S$-type coupling is
\begin{equation}
  {\Delta}  \bar A^{(2)} _{S\pm}(r,y_{\pm}) 
  =  {\Delta}  \left[A^{(2)} _{S\pm}(r,y_{\pm}) 
       -Y^{(2)} _{S\pm}(r,y_{\pm}) \right], 
\end{equation}
with $A^{(2)} _{S}(r,y_{\pm})$ given by Eq.
(\ref{eq:A_[GGK] at y_+-}), and $Y^{(2)} _{S}(r,y_{\pm})$ by Eq. (\ref{eq:Y^J_[GGS] y_+-}). Once we know $\bar A^{(2)}$ and $Y^{(2)}_{S}$, the spatial component of the metric perturbations $\bar B^{(2)}$ is obtained from Eqs.~(\ref{eq:bar B^2 rew 1}), (\ref{eq:Y^J_[GG0] y_+-}), and (\ref{eq:Y^J_[GGS] y_+-}) as 
\begin{eqnarray}
   \Delta \bar B_{\pm}^{(2)} (r,y_{\pm})
   &=& - \frac{1}{2} \Delta \bar A_{\pm}^{(2)}  
 + 4\pi G_\pm a^2_\pm T^{(2)}_\pm
 - \Delta S_{B\pm} 
\nonumber
\\ 
&& 
+ 3N  \biggl[
      \pm a_{\pm}^4
     S_{\xi}^{\pm} 
    + \int^{y_{-}}_{y_{+}} dy 
     \Bigl(
  a^2 v_{\pm} \Delta {\mathbb S}_{\varphi\pm} 
    + \frac{3u_{\pm}} {2\kappa \dot \phi_0^2} \Delta 
      {\mathbb S}_{Y\pm} -a^2 S_{\psi\pm} 
  \Bigr)
\biggr] 
- \frac{3}{2} \Delta Y^{(2)}_{S\pm}
 \,.
\label{eq:bar B^2 rew}
\end{eqnarray}

To identify the order of magnitude of various terms in the second order perturbations, we have to keep track of the powers of both $r_\star$ and $a_-$. 
Terms with additional inverse powers of $r_\star$ are basically suppressed for long wavelength perturbations. However, as we have seen for the KK mode and the short-ranged part in the analysis of the linear perturbations, a complication arises due to the existence of a large nondimensional hierarchy factor $1/a_-$. Here we continue to use the convention $a_+ =1$. The dependences of the perturbation variables on the warp factor and on $r_\star$ are summarized as 
\begin{eqnarray}
A_{0\pm}^{(1)} \sim a^0, ~~~~
Y_{0\pm}^{(1)} \,, ~ \varphi_{0\pm}^{(1)}  \sim \frac{a_{\mp}^2}{a^2}+1, ~~~~
\hat\xi^y_{\pm} \sim \frac{1}{a_{\pm}^2} \,,
\label{eq:count a r 1}
\end{eqnarray}
and 
\begin{eqnarray}
A_{S+}^{(1)} \sim \frac{1}{ a^2 r_\star^2},
 ~~~~
A_{S-}^{(1)} \sim \frac{1}{ a^4 r_\star^2},
 ~~~~
Y_{S\pm}^{(1)} \,, ~ \varphi_{S\pm}^{(1)} \sim \frac{1}{a^2_{\pm} a^2 r_\star^2}.
\label{eq:count a r 2}
\end{eqnarray}

In the following subsection, we first evaluate the terms from the zero-type coupling, classifying them into three parts: the part to recover the four-dimensional Einstein gravity, the manifestly suppressed corrections, and the unsuppressed corrections. 
The unsuppressed correction is later shown to be canceled by the contribution from the terms of $S$-type. 
We stress that a weak back reaction is assumed only when we roughly estimate the dependence on $r_\star$ and $a_-$.

        \subsection{ $\bar A_{0}$}

Let us consider $\bar A_{0}$ given in Eq.~(\ref{eq:A_[GG0] rew}).  
From Eqs.~(\ref{eq:S_Sigma}) and (\ref{eq:S_xi}) with the large coupling limit (\ref{eq:varphi^J at y_pm}), $(S_{\xi}^{\pm}- S_{\Sigma}^{\pm})$ becomes 
\begin{eqnarray}
  a^4_\pm (S_{\xi}^{\pm}- S_{\Sigma}^{\pm})
 &=& 
  a^2_\pm \biggl[
   \underline{ 
    3 a^2_{\pm} \int dr B_{,y}Y_{,r}
  - 3 \int dr \hat \xi^y_{,r} \Delta Y
   } 
 +   \underline{\underline{
      a^2_{\pm} A_{,y}(Y-4H \hat\xi^y)
 +  \hat \xi^y \Delta Y 
 -  2H (\hat \xi^y_{,r})^2 
     }} 
\biggr]
\nonumber
\\
& &
 + a^2_\pm \biggl[
  (2\hat \xi^y_{,r} \partial_r - \Delta \hat \xi^y)
  (Y+2H \hat \xi^y)
+
  B \Delta \hat \xi^y - \hat \xi^y \Delta A
 + \hat \xi^y_{,r} (B-A)_{,r}
+
   \hat\xi^r ( a^2_{\pm} A_{,y} 
    + \Delta \hat \xi^y)_{,r} \biggr]
\\
&& 
 - a^4_{\pm} 
   \int \Bigl[ 
      2B_{,y}A_{,r} + \frac{r}{4} B_{,ry} (B+3A)_{,r}
   \Bigr] dr
  \,.
   \nonumber
\end{eqnarray} 
As for ${\mathbb S}_\varphi$, the last two terms in Eq. 
(\ref{eq:Def. mathbb S_varphi}) are rewritten as
\begin{eqnarray}
  (a^2 v_{\pm})
  \Bigl(   H Y^2 - \frac{\kappa}{3} \partial_y       
        \bigl[ \varphi^2 \bigr]
  \Bigr) 
  =  
  \partial_y 
  \Bigl(
     \underline{\underline{ 
        \frac{ a^2}{2} \frac{\varphi}{\dot \phi_0} Y 
    }}
  -  \underline{\frac{\kappa a^2}{3} \varphi^2 v_{\pm}
}   \Bigr)
  +  \frac{3 Y \Delta Y}{4 \kappa \dot \phi_0^2 }
  - \underline{\frac{1}{2} a^2 u_{\pm} Y^2} \,.
\label{eq:S_phi rewrite}
\end{eqnarray}

The expression (\ref{eq:A_[GG0] rew}) starts with terms of $\sO(1/r_\star^2)$; hence we start our discussion with these leading order terms. Here, to understand the absence of terms of $\sO(r_\star^{0})$, we need to notice that $\partial_y A^{(1)} $ and $\partial_y B^{(1)} $ do not have contributions from the zero mode, and hence they are $\sO(1/r_\star^2)$. 
Let us identify the dependence on the hierarchy $a_-$ of each term, concentrating on the case that the matter fields are on the negative tension brane. 
For this purpose, we can use Eq.~(\ref{eq:count a r 1}). As for $\partial_y A^{(1)}$ and $\partial_y B^{(1)}$, we use Eq. (\ref{eq:count a r 2}) instead because the zero mode contribution exactly vanishes. The terms in the second line of Eq.~(\ref{eq:A_[GG0] rew}) possess a $y$ integration. 
This integration is basically dominated by the contribution from the neighborhood of the negative tension brane. One exception is the case in which the integrand has the quadratic form of the zero mode contribution of the TT variables ($A_0$ and $B_0$ multiplied by $a^2$ as $a^2 A_0\times A_0$). 
This integration does not have any inverse powers of $a_-$. 
The other exception is the case in which the integrand contains the factor of ${u_-/\kappa\dot\phi^2_0}$. 
The formulas for this case are summarized in Eqs.~(\ref{eq:approx int}) and we find that only the terms with the integrand proportional to 
${u_-^3/\kappa\dot\phi^2_0}$ give a correction that behaves as $1/a_-^6$. The other terms are at most $\sO(1/a_-^4)$. Hence, we can pick up the terms with a large power of $1/a_-$ just by looking at the behavior of the integrand near the negative tension brane. 
Then, we find that, among the terms of $\sO(1/r_\star^2)$, the terms associated with a single underline or with double underlines behave as $1/a_-^4$ or $1/a_-^2$. Since the usual post-Newtonian correction in the four-dimensional Einstein gravity is of $\sO(a_-^{0}/ r_\star^2)$,  we expect that the terms with underlines cancel each other, and we show that, in fact, this is the case. 
The terms with a single underline completely cancel each other. 
For example, the term $a^2 \dot H Y^2$ in ${\mathbb S}_Y$ is canceled with the last term of Eq.~(\ref{eq:S_phi rewrite}). Here, it is worth mentioning that the cancellation occurs separately within the terms of different types:  the terms quadratic in the TT variables, those bilinear in the TT variables and the scalar-type variables ($Y$, $\varphi$, and $\xi^y$), and those quadratic in the scalar-type variables. 
The terms with double underlines do not vanish completely, but they are, in total, combined to terms of $\sO(a_-^0/r_\star^2)$ with the aid of Eqs.~(\ref{eq:comb Y varphi}) and (\ref{eq:comb Y 2H xi}) when we consider the long-ranged part. 
The contributions from the short-ranged part cannot be combined to reduce the power of the warp factor, but they are at most $\sO(\ell^2 /a_-^4 r_\star^4)$.
After a straightforward calculation, the remaining terms give the usual post-Newtonian term in four-dimensional Einstein gravity (Appendix \ref{sec:evaluation}). 
This result also applies for the case that the matter field is on the positive tension brane because any term of $\sO(1/r_\star^2)$ irrespective of the power of $a_-$ was not discarded in the above computation; namely, we obtain in the isotropic gauge (\ref{eq:iso gauge}) 
\begin{eqnarray}
 {\Delta}   
   \bar A^{(2)}_{0\pm} (r,y_{\pm})
 &=& 
   8 \pi G 
   (\rho^{(2)}_{\pm} + 3 P^{(2)}_{\pm} )
   - 4 \Phi_\pm  {\Delta}  \Phi_\pm 
   + \sO\left({\ell^2 \over r_\star^4}\right)\,.   
\label{eq:sol bar A^2_[000]}
\end{eqnarray}

Next, we consider the terms of $\sO(1/r_\star^4)$. 
Again, we begin by considering the case that the matter fields are on the negative tension brane. As we did for the terms of $\sO(1/r_\star^2)$, we can identify the dependence on hierarchy $a_-$ of these terms using the estimates (\ref{eq:count a r 1}), (\ref{eq:count a r 2}), and (\ref{eq:approx int}). 
Then, the terms with the highest inverse power of $a_-$ start with $1/a_-^6$, which we refer to as $F$ terms. They are given by 
\begin{eqnarray}
   F_\pm &:=& 
   -  \Delta \bigl( \hat\xi_{ \pm }^y \bar{A} _{,y} \bigr)
   + F_{ {\mathbb S}_\varphi, {\mathbb S}_Y \pm } \,,
\label{eq:def. F}
\\
   F_{ {\mathbb S}_\varphi, {\mathbb S}_Y \pm } 
&=&   2 N  \Delta \int dy dr \,
  \biggl\{ 
     \frac{3u_{\pm}} {2\kappa \dot \phi_0^2} 
   \Bigl[
    \Bigl( \frac{2 \kappa}{3} 
        \varphi \partial_r \Delta \varphi 
 -3 Y_{,r}  \Delta Y  
    \Bigr)
   + Y(\Delta Y)_{,r} 
   + 2 a^4 B_{S,y} \Bigl( \frac{Y_{,r}}{a^2} \Bigr)_{,y}
   \Bigr] 
      - a^2 v_{\pm} B_{S,y}  Y_{,r} 
    \biggr\}    \,,
\nonumber
\end{eqnarray}
where we have taken into account 
Eqs.~(\ref{eq:comb Y varphi}) and (\ref{eq:comb Y 2H xi}). 
The first term in the curly brackets in  
$F_{{\mathbb S}_\varphi, {\mathbb S}_Y \pm}$ comes from 
${\mathbb S}_Y$ and the second term from ${\mathbb S}_\varphi$. 
The remaining terms are at most $\sO(\ell^2/a_-^4 r_\star^4)$. 
The relative amplitude of these remaining terms compared to the ordinary post-Newtonian corrections is $\sO(\beta_-)$ or $\sO(\gamma_-)$.  
Therefore, only the $F$ terms have the possibility of introducing a non-negligible correction. 
However, we will show in the succeeding subsection that the contribution from the $F$-terms is also completely canceled by that from couplings of the $S$-type.

Now we consider the case that the matter fields are on the positive tension brane.  
In counting the order of each term with respect to $a_-$, we will notice that the inverse power of $a_-$ can appear only from a contribution near the negative tension brane.  
Furthermore, from the estimates (\ref{eq:count a r 1}) and 
(\ref{eq:count a r 2}), we find that the variables in the first order perturbation are at most of $\sO(1/a_-^2)$, and such enhanced variables are associated with the factor $1/r_\star^2$. 
With this notion and the estimate (\ref{eq:approx int}), it will be easy to verify that all the terms quartic in $1/r_\star$ in $\Delta \bar A^{(2)}_{0+}$ are, at most, 
$\sO(\ell^2/a_-^2 r_\star^4)$; namely, they are suppressed compared to the ordinary post-Newtonian corrections by the factor of $\sO(\beta_+/a_-^2)$ or $\sO(\gamma_+/a_-^2)$.

The suppression factors that we encounter at the second order are not as small as those in the linear perturbation, $\beta_+$ and $\gamma_+$.  
This is a natural consequence of our approximation of gradient expansion. 
Near the negative tension brane, the conditions that the scale of the spatial gradient is larger than the typical length scales $\ell$ and $m_S^{-1}$, respectively, become $(\beta_+/a_-^2)= (\ell^2/a_-^2 r_\star^2)\ll 1$ and  
$(\gamma_+/a_-^2)= (1/\tilde m_S^2 a_-^2 r_\star^2)\ll 1$.  
Although the correction seems to become large when we consider the case with large $1/a_-$, we think that this is an artifact due to the limitation of the present approximation. When we do not have a bulk scalar field, it has been proved that the correction to the four-dimensional Einstein gravity in the $(1/a_-)\to \infty$ limit stays small\cite{Garriga:2000yh}. 

\subsection{  $A_S$ and $Y_S$ }

In this subsection, we discuss the terms $A^{(2)}_{S}$ and $Y^{(2)}_{S}$. The contribution of these terms completely cancels the correction due to the $F$ terms. 

From Eq.~(\ref{eq:KK mode A^J}), we obtain 
\begin{eqnarray}
  A^{(2)} _{S\pm}(r,y_{\pm})
    &=&
     4 N^2   
       a_{\pm}^4 
       \Bigl[ \kappa \Bigl(P_{\pm}^{(2)}+ \frac{2}{3} \rho_{\pm}^{(2)}\Bigr)
       \pm S_{\Sigma}^\pm ) \Bigr]
      \int^{y_{-}}_{ y_{+} } v_{\pm}^2 dy' 
\nonumber
\\
 && + 2N 
   \int^{y_{-}}_{y_{+}} dy' \frac{v_{\pm}}{a^2 }
       \int^{y'}_{y_{\mp}} dy'' a^2 S_{A\pm}  
      - (2N)^2
       \Bigl( \int^{y_{-}}_{y_{+}} v_{\pm}^2 dy\Bigr)
           \int^{y_{-}}_{y_{+}} a^2 S_{A\pm} dy \,,
\label{eq:A_[GGK] rew}
\end{eqnarray}
where we have performed an integration by parts by using Eqs.~(\ref{eq:dot u v}), and also we have used again the fact that $S_\Sigma$ on the vacant brane is zero, as well as Eqs.~(\ref{eq:v+ and v- relation}) and (\ref{eq:u v on y+-}).

First, we concentrate on the case with the matter fields on the negative tension brane. 
The first term in the square brackets in Eq. (\ref{eq:A_[GGK] rew}) is suppressed by a factor of $\ell^2/r_\star^2$ compared with the first term on the right hand side in Eq.~(\ref{eq:sol bar A^2_[000]}), and hence can be neglected. The other terms in 
$\Delta A^{(2)} _{S-}$ are quartic in $1/r_\star$ or smaller.  Hence, we have only to study the terms that give a correction of $\sO(\ell^2/a_-^6 r_\star^4)$. Neglecting the terms higher order in $1/r_\star$, the contribution from $S_\Sigma^-$ becomes
\begin{eqnarray}
   - 4N^2 a^2_- \Bigl( \int v^2_- dy\Bigr)
    \Biggl(- \int B _{-,y} Y _{-,r} dr
    - 3 Y_- A_{-,y}
  + \frac{2}{a_-^4}  
  \int \frac{1}{r} Y_{-,r} \hat\xi^y_{-,r} {dr}
   \Biggr) 
 + \sO \Bigl( \frac{\ell^2}{a_-^4 r^2_\star} \Bigr) 
\,,
\label{eq:S_sigma leading}
\end{eqnarray}
where we dropped the terms proportional to $Y+2H\hat\xi^y$ because Eq.~(\ref{eq:comb Y 2H xi}) shows that this combination becomes higher order in $1/a_-$. 
As for the terms containing $S_A$ in Eq.~(\ref{eq:A_[GGK] rew}), the contribution of $\sO(\ell^2/(a_-^6 r^4_\star))$ comes from the terms with underlines in Eq. (\ref{eq:S_A full exp}).  For these terms, the integral of $a^2 S_{A-}$ can be performed explicitly as 
\begin{eqnarray}
 \int_{y_+}^y a^2 S_{A-}dy = a^2 
       \Bigl(
      3a^2 Y_- A_{-,y} + \int dr a^2 B_{-,y} Y_{-,r}
      - 2 \int \frac{dr}{r} Y_{-,r} 
         \frac{\varphi_{-,r}}{\dot\phi_0}
      \Bigr)\,.  
\label{eq:intSA}
\end{eqnarray}
Here, note that the contribution from the boundary at $y=y_+$ vanishes.  Using Eq. (\ref{eq:intSA}), we find that the last term in Eq.~(\ref{eq:A_[GGK] rew}) cancels the leading order contribution from $S_\Sigma^-$ of Eq. (\ref{eq:S_sigma leading}). 
Then, the remaining parts in Eq.~(\ref{eq:A_[GGK] rew}) give 
\begin{eqnarray}
 \Delta A^{(2)}_{S- } (r,y_{-})
     = 2N \Delta \int^{y_{-}}_{y_{+}}
     dy v_- 
    \Bigl(
      3a^2 Y_- A_{-,y} + \int dr a^2 B_{-,y} Y_{-,r}
       - 2\int \frac{dr}{r} Y_{-,r} 
          \frac{\varphi_{-,r}}{\dot\phi_0}
      \Bigr)
   + \sO \biggl(\frac{\ell^2}{a_-^4 r^4_\star} \biggr).
 \label{eq:A[00K]}
\end{eqnarray}

The other correction that we have not considered yet comes from $Y^{(2)}_{S-}$. To evaluate the expression presented in Eq.~(\ref{eq:Y^J_[GGS] y_+-}),  first we need to evaluate $\Delta Y_{0}$ given in Eq.~(\ref{eq:zero mode of Y^J(r,y)}). 
Only the leading terms of $O(1/a_-^{4}r_\star^2)$ in $\Delta Y_{0}$  are relevant, and they are evaluated as 
\begin{eqnarray}
 - \Delta Y^{(2)}_{0-}(r,y)
&\approx& 
   \int dr \Bigl[
   \frac{3}{2r^{8/3}} \partial_r
\Bigl\{ r^{8/3} 
    \Bigl[(Y_{-,r})^2+ \frac{2\kappa}{3}(\varphi_{-,r})^2 
    \Bigr]    
\Bigr\}   -2a^2 B_{-,y} Y_{-,yr} 
\Bigr]
\nonumber
\\
&& 
  + 2N a^2\sum_\sigma \sigma u_\sigma (y) 
    \biggl\{ 
    \Delta 
  \Bigl(
         \frac{Y_- \varphi_- }{ 2\dot\phi_0^2} 
     - \frac{\kappa \varphi_-^2}{3}v_{-\sigma}(y)
 \Bigr)
\nonumber
\\
&& 
+   \int dr \Bigl[
      2a^2 Y_{-,r} B_{-,y}
      - \frac{3}{2r^{8/3}} \partial_r
    \Bigl( 
      r^{8/3} Y_{-,r} \frac{\varphi_{-,r}}{ \dot\phi_0^2} 
    \Bigr)
    \Bigr] 
\biggr\}. 
\label{eq:Delta Y}
\end{eqnarray}
Note that the terms from $S_{\xi}^\sigma$ in Eq.~(\ref{eq:zero mode of Y^J(r,y)}) cancel the terms obtained by setting $y=y_\sigma$ after the $y$ integration in Eq.~(\ref{eq:zero mode of Y^J(r,y)}). Substituting Eq. (\ref{eq:Delta Y}) into Eq. (\ref{eq:Y^J_[GGS] y_+-}), we obtain 
\begin{eqnarray}
- \Delta Y^{(2)}_{S-}(r,y_{-})
  &=&
  {3N\Delta}\Biggl\{ 
    \int^{y_{-}}_{y_{+}} dy dr
     \frac{u_{-}}{\kappa \dot \phi_0^2} 
    \Bigl[  
    \Bigl( 3 Y_{-,r}\Delta Y_- -\frac{2 \kappa}{3} \varphi_- \partial_r 
    \Delta \varphi_- \Bigr)
   + \frac{4\kappa}{3r}  (\varphi_{-,r})^2 
   -2 a^4 B_{-,y} \Bigl( \frac{Y_{-,r}}{a^2} \Bigr)_{,y}
   - Y_- \partial_r \Delta Y_-
   \Bigr]
\nonumber
\\ 
&&  +
   \int^{y_{-}}_{y_{+}}dy dr
     \frac{u_{-}} {\kappa \dot \phi_0^2} 
    \Bigl[  
     Y_- \partial_r \Delta Y_-
   +  \frac{H}{\dot\phi_0}
     \Bigl(
        \varphi_- (\Delta Y_-)_{,r} + Y_-(\Delta \varphi_-)_{,r}
     \Bigr)
- \frac{2}{r}  Y_{-,r}
   \Bigl( Y_{-,r}  + \frac{2 H \varphi_{-,r}}{\dot \phi_0} 
   \Bigr)
   \Bigr]\Biggr\}
\nonumber
\\
&&
   + O\Bigl( {\ell^2 \over a_-^4 r_\star^4} \Bigr)  \,.
\label{eq:Y[GGS]}
\end{eqnarray}
From Eq.~(\ref{eq:comb Y varphi}) the terms inside the second integral turn out to be $O(\ell^2/a_-^4 r_\star^4)$.

Combining Eqs. (\ref{eq:def. F}), (\ref{eq:A[00K]}), and (\ref{eq:Y[GGS]}), we obtain
\begin{eqnarray}
F_- + \Delta(A_{S-}^{(2)} - Y_{S-}^{(2)})  
&=&
   \Delta\left\{
 - \hat\xi_{-}^y \bar{A} _{-,y} 
 + 4N  \int dy
\biggl[u_{-} \int  
  \frac{dr}{r} \frac{(\varphi_{-,r} )^2}{\dot \phi_0^2} 
 + 
  v_{-} \Bigl( \frac{3}{2} a^2 Y_- A_{ -,y}- \int 
  \frac{dr}{r} Y_{-,r} \frac{\varphi_{-,r} }{\dot\phi_0 } 
\Bigr)
\biggr]
\right\}
\nonumber
\\
&&
 + \sO\Bigl( \frac{ \ell^2 }{a_-^4 r_\star^4} \Bigr) .
\label{eq:F+A-Y}
\end{eqnarray}
After writing the above expression in terms of $u_-$ and $v_-$, we can perform the integration with respect to $y$ by using Eq.~(\ref{eq:dot u v}). 
Then, with the aid of Eq.~(\ref{eq:u v on y+-}), we find that Eq. (\ref{eq:F+A-Y}) reduce to terms higher order in $1/r_\star^2$ or those of $\sO (\ell^2/a_-^4 r_\star^4)$. 
Hence, our conclusion is 
\begin{eqnarray}
  F_- + \Delta( A_{S-}^{(2)}-Y_{S-}^{(2)}) = 
    \sO\left( {\ell^2 \over a_-^4 r_\star^4}\right). 
\end{eqnarray}
The weak back reaction was assumed only for evaluating the order of the residual terms. 

In the case with the matter fields on the positive tension brane, the corrections both from $A^{(2)} _{S+}$ and from $Y^{(2)} _{S+}$ are suppressed by either $(\beta_+/a_-^2)$ or $(\gamma_+/a_-^2)$ for the same reason as before. 

\subsection{ Spatial components of TT part}

Now the evaluation of $\bar B^{(2)}$ in the isotropic gauge is straightforward. 
Substituting the first order quantities and the result of $\bar A^{(2)}$ into Eq.~(\ref{eq:bar B^2 rew}), we basically obtain 
\begin{eqnarray}
 \Delta
  \bar B^{(2)} (r,y_{\pm})
 &=&
   - 8 \pi G \rho^{(2)}_{\pm} 
 + 4 \Phi_\pm \Delta \Phi_\pm   
-  \left( \Phi_{\pm,r} \right)^2
 +\cdots\,, 
\label{eq:sol bar B^2_[GG0]'}  
\end{eqnarray}
which is identical to the result for the four-dimensional Einstein gravity in isotropic coordinates except for the residual denoted by $(\cdots)$ \cite{Kudoh:2001wb}.

In the case with the matter field on the negative tension brane, these residual terms are 
\begin{equation}
    F_{S_B-} 
 + \frac{3}{2} 
   (F_{ {\mathbb S}_\varphi, {\mathbb S}_Y-} 
 - \Delta Y^{(2)}_{S-}) 
 + \sO \Bigl( \frac{\ell^2 }{a_-^4 r_\star^4} \Bigr) \,,
\label{eq:sol bar B^2_[GG0]}  
\end{equation}
where we introduced 
\begin{eqnarray}
F_{S_B\pm} &=&  
   \frac{1}{9N^2 a^6_\pm} 
   \Delta \left[ \int \frac{dr }{r} 
   (\Phi_{\pm,r})^2 \right] \,.
\label{eq:F_SB}
\end{eqnarray}
In the same way as for $\bar A^{(2)}$, cancellation occurs for the leading order in $1/a_-$ as 
\begin{eqnarray}
  \mathrm{Eq.~} (\ref{eq:sol bar B^2_[GG0]}) 
   &=&   \Delta\left\{ F_{S_B-}
       + 3N  \int^{y_-}_{y_+} dy dr 
\biggl[  
 \frac{2u_{-} } { \dot \phi_0^2} 
   \frac{(\varphi_{-,r} )^2}{r}
- a^2 v_{-} B_{-,y}Y_{-,r}
\biggr]\right\}
 + \sO \left({\ell^2 \over a_-^4 r_\star^4}\right) \,,
\cr
&=& \sO \left({\ell^2 \over a_-^4 r_\star^4}\right)\,. 
\end{eqnarray}

In the case with the matter fields on the positive tension brane, the residual terms represented by $(\cdots)$ in Eq.~(\ref{eq:sol bar B^2_[GG0]'}) are, at most, $O(\ell^2/a_-^2 r_\star^4)$ as before. 
To conclude, the four-dimensional Einstein gravity is approximately recovered under the assumption of the large coupling limit. The corrections to four-dimensional Einstein gravity are suppressed by the factor of $\sO (\beta_{\pm}/a_{\mp}^2)$ or $\sO (\gamma_{\pm}/a_{\mp}^2)$.

    \section{Summary}
    \label{sec:summary}

In this paper, we have considered the second order gravitational perturbations in the RS two branes model with the radius stabilization mechanism. As a model for the radius stabilization, we have assumed a scalar field that has a potential in the bulk and a potential on the brane. 
From the five-dimensional Einstein equations, the master equations for the TT part of the metric perturbations and for the scalar-type perturbation are derived assuming static axisymmetric configurations. 
We have presented formal solutions of these equations by means of the Green's function. 
We have shown an iterative scheme to obtain approximate solutions by applying the derivative expansion method for the massive modes.  
For the validity of the derivative expansion, the physical mass of the massive modes must be sufficiently large on the respective branes.  
This sets a limitation on our scheme. 
Taking infinite separation distance between two branes is beyond the framework of the present analysis because the mass of the lowest KK mode becomes zero.

We have shown the recovery of the four-dimensional Einstein gravity in the second order perturbations in the following limit:
(1) The coupling between the scalar field and the branes is infinitely large 
[see Eq.~(\ref{eq:large limit})].
(2) We consider the perturbations induced by the matter fields on one brane where we reside, and neglect the effects caused by the matter fields on the other brane [see Eq.~(\ref{eq:simp})].

When we consider the case in which the matter fields are on the negative tension brane, the correction to the four-dimensional Einstein gravity appears at the relative order of $\sO((a_+/a_-)^{4}(\ell/r_\star)^2)$, where $\ell$ is the AdS curvature scale, $r_\star$ is the typical length scale of the perturbation, and $(a_+/a_-)$ is the ratio between the warp factors on the positive and the negative tension branes. 
When this ratio $(a_+/a_-)$ is $\sO(10^{16})$, the hierarchy between Planck and electro-weak scales can be explained. 
With this choice of the hierarchy, the correction to the metric in the linear perturbation becomes comparable to the usual Newtonian potential when 
$r_\star\lesssim 0.1$mm. However, this correction does not give a contribution to the force outside the matter distribution. 
Hence, it seems to be harmless in reproducing the predictions of four-dimensional Einstein gravity. 
We have not confirmed if this feature remains in the second order perturbation, but the correction is suppressed by the above factor compared to the usual post-Newtonian correction. 
Hence, the effect due to this correction is almost impossible to detect.

When we consider the case in which the matter fields are on the positive tension brane, the correction to the four-dimensional Einstein gravity in the linear perturbation appears at the relative order of $\sO((\ell/r_\star)^2)$, while the correction in the second order perturbation is 
$\sO((a_+/a_-)^2 (\ell/r_\star)^2)$ compared to the usual post-Newtonian terms. 
Hence, it seems that the deviation from four-dimensional 
Einstein gravity appears at a larger scale in the second order perturbation. 
However, this is very likely to be an artifact due to the limitation of our approximation scheme.

To give a complete proof of the recovery of the four-dimensional Einstein gravity, further extension of the present analysis will be necessary. Here we considered the large coupling limit.  It will be interesting to evaluate the dependence of the correction on the coupling strength. 
Furthermore, to take into account the contributions from the matter fields on the other brane will be interesting. 
To investigate these issues, a formulation along the line of this paper will be promising. Through this second order calculation, we have encountered many miraculous cancellations. 
This might be due to our possibly bad choice of gauge. 
We would like to defer pursuing a more simplified derivation to a future publication, in which we will discuss the unsolved issues mentioned above.

\vspace{0.5cm}
    \centerline{\bf Acknowledgments}

We would like to thank Takashi Nakamura and Hideo Kodama for informative comments. T.T. would like to thank IFAE, where this work was partly done, for cordial hospitality. This work was supported by the Monbukagakusho Grant-in-Aid No.~1270154 and the Yamada Foundation.

 \appendix

\section{4D Einstein gravity}

We have used the isotropic gauge (\ref{eq:iso gauge}) to fix the radial gauge coordinates because it is easy to compare with the four-dimensional Einstein gravity. However the calculation of the second order perturbation becomes slightly easier by taking $\xi^r=0$, although we do not previously know the corresponding four-dimensional Einstein gravity.
In this appendix we derive the expression for the result of the metric perturbations in four-dimensional Einstein gravity in an arbitrary choice of the radial gauge, which corresponds to the various choice of
$\hat\xi^r$ in Eq.~(\ref{eq:gauge trans ABC}).

In general, the radial gauge transformation from the isotropic gauge ${\bar{A}} _{IS}$ to an arbitrary gauge ${\bar{A}} _{\xi^r}$ is given by 
\begin{eqnarray}
    {\bar{A}}^{(2)}_{\xi^r} = 
    {\bar{A}}^{(2)}_{IS} 
   - \zeta^r {\bar{A}}^{(1)}_{IS,r} \,,
\label{A1}  
\end{eqnarray}
where the generator $\zeta^r$ is related to the quantities at the first order perturbation by the gauge transformation law, 
\begin{eqnarray}
     {\bar{A}}^{(1)}_{\xi^r}-{\bar{B}}^{(1)}_{\xi^r} 
  =  {\bar{A}}^{(1)}_{IS}-{\bar{B}}^{(1)}_{IS} 
      + 2\zeta^r_{,r} \,. 
\end{eqnarray}
On the other hand, these bared quantities are also related to the quantities in the Newton gauge as 
\begin{eqnarray}
   \bar A_{\xi^r}^{(1)} - \bar B_{\xi^r}^{(1)} -2 \xi^r_{,r}   
  =  A ^{(1)} - B^{(1)} 
  =  \bar A_{IS} ^{(1)} - \bar B_{IS}^{(1)} 
   - 2\xi^r_{IS,r}  \,,
\end{eqnarray}
where $\xi^r_{IS}$ is defined by Eq.~(\ref{eq:iso gauge}). Therefore, we find that $\zeta^r$ is simply given by $\zeta^r=\xi^r-\xi^r_{IS}$. Substituting this relation into Eq.~(\ref{A1}), we obtain 
\begin{eqnarray}
    {\bar{A}}^{(2)}_{\xi} 
  = {\bar{A}}^{(2)}_{IS} 
   +{\bar{A}}^{(1)}_{IS,r}  
    ( \xi^r_{IS} - \xi^r )  \,.
\label{eq:4D Ein. for xi^r} 
\end{eqnarray}
By this equation, the metric perturbation of the four-dimensional Einstein gravity in an arbitrary gauge $\xi^r$ is determined.

\section{Useful formulas}

In the calculation of the second order perturbations, we often use some relations and results that are easily derived from the original definitions and equations, but we have not derived them explicitly. It is convenient to summarize such results, and so we devote this appendix to giving the useful relations and formulas.

\subsection{$u_\pm$ and $v_\pm$}

We gives some properties of the functions $u_\pm$ and $v_\pm$ which are defined by Eqs.~(\ref{eq:def. v}) and (\ref{eq:def. u_+-}).
The differentiation of these functions with respect to $y$ are 
\begin{eqnarray}
  \partial_y (a^2 u_\pm) = -2 a^2\dot H v_\pm \,,  
\quad
  \partial_y (a^2 v_\pm)= a^2  \,,  
\quad 
  \partial_y v_\pm = u_\pm   \,. 
\label{eq:dot u v}
\end{eqnarray}
The last equation is particularly useful to integrate ${\mathbb S}_\varphi$ and ${\mathbb S}_Y$ with respect to $y$. 
From the definition, $v_+$ and $v_-$ are related as 
\begin{eqnarray}
    v_- -v_+ = \frac{1}{2Na^2}\,.
    \label{eq:v+ and v- relation}
\end{eqnarray}
On the branes 
$u_\pm$ and $v_\pm$ become 
\begin{eqnarray}
    u_\pm(y_\pm) &=&1 \pm \frac{H(y_{\pm})}{a^2_{\pm}N} \,,
\quad
    u_\pm(y_\mp) = 1 \,,
\nonumber
\\
   v_\pm(y_\pm) &=& \mp \frac{1}{2a^2_{\pm}N} \,,
    \qquad
   v_\pm(y_\mp) = 0 \,.
\label{eq:u v on y+-}
\end{eqnarray}
From Eqs.~(\ref{eq:dot u v}) and (\ref{eq:u v on y+-}), we obtain 
\begin{eqnarray}
  \int^{y_-}_{y_+} dy~  u_\pm v_\pm^2 = \frac{1}{24N^3 a_\pm^6}\,.
\end{eqnarray}
The sums of $+$ and $-$ modes are 
\begin{eqnarray}
\sum_\sigma \sigma u_{\sigma}(y)&=& \frac{H}{N a^2} \,,
\qquad
\sum_\sigma \sigma v_{\sigma}(y) =  - \frac{1}{2N a^2} \,,
\cr
  \sum_\sigma \sigma u_{\sigma}(y)u_{-\sigma}(y) 
&=&  0  \,,
\qquad
    \sum_\sigma \sigma u_{\sigma}(y)v_{-\sigma}(y)
    = \frac{1}{2Na^2}   \,,
\label{eq:sum v u etc}
\end{eqnarray}
and also
\begin{eqnarray}
   \sum_{\sigma=\pm} u_{\sigma}(y_{\pm}) f(y_{\sigma})
   &=& \pm \frac{H}{a_\pm^2 N}  f(y_{\pm})  
       + \sum_{\sigma=\pm} f(y_{\sigma}) \,,
\cr
  N \sum_{\sigma=\pm}  \sigma a^2_{\sigma} u_{\sigma}(y_{\pm})
      f(y_{\sigma})
   &=& H(y_{\pm}) f(y_{\pm})  
     +  N \sum_{\sigma=\pm}  \sigma a^2_{\sigma}
     f(y_{\sigma}) \,,
\cr
  \sum_\sigma \sigma v_\pm (y_{\sigma}) f (y_{\sigma})
   &=&  - \frac{f (y_{\pm})}{2N a_{\pm}^2} \,,
\cr
  u_{\pm} (y_{\sigma}) &=& u_{\sigma} (y_{\pm}) \,. 
\label{eq:rel-Sum u}
\end{eqnarray}

\subsection{Equations}
\label{appen:equations}

From Eqs.~(\ref{eq:Ein-A^J}) and (\ref{eq:gauge trans ABC}) the derivatives of $\bar A^{(1)}$ with respect to $y$ are  
\begin{eqnarray}
   &&  \bar A_{,y}^{(1)} (r,y) =  A_{,y} -\frac{2\kappa}{3} \dot \phi_0
    (\varphi  - \dot \phi_0 \xi^y ) \,,
\\
  && \bar A_{,yy}^{(1)} (r,y) 
    =   \frac{\Delta}{a^2} (Y-A) 
    - 4H A_{,y} 
    - \frac{4 \kappa}{3} \ddot \phi_0
(\varphi - \dot \phi_0 \xi^y ) \,,
\end{eqnarray}
and using this result we obtain 
\begin{eqnarray}
  && \partial_y[ \xi^y \bar A_{,y} + \xi^r \bar A_{,r}]
    =  A_{,y}(Y-4H\xi^y)
    +  \xi^y \frac{\Delta}{a^2} (Y-A) 
   - \frac{1}{a^2}\xi^y_{,r} \bar A_{,r}
   +  \xi^r \bar A_{,ry} 
   - \frac{2\kappa}{3} ( \dot\phi_0 Y +2\ddot\phi_0 \xi^y)
    (\varphi- \dot\phi_0 \xi^y)
    ~\,.
\end{eqnarray}
We have often evaluated $A_{,y}$ and $B_{,y}$ on the brane, which are determined by the junction condition (\ref{eq:jun. for A_,y}).  
These quantities are purely KK mode contributions. 
Equation (\ref{eq:KK mode A^J}) gives  
\begin{eqnarray}
    A^{(1)}_{S,y}(r,y) 
    &=& - \frac{8}{3 a^2} \sum_\sigma v_\sigma \Delta \Phi_{\sigma}
\,, ~~
    B^{(1)}_{S,y}(r,y) 
    = \frac{8}{3r a^2} \sum_\sigma 
      v_\sigma  \partial_r \Phi_{\sigma}  \,,
\label{eq:A_S explicit}
\end{eqnarray}
and, taking $y=y_\pm$ ,
\begin{eqnarray}
    A^{(1)}_{S,y}(r,y_{\pm}) 
     &=& \pm \frac{4}{3N a_{\pm}^4} \Delta \Phi_{\pm}
\,, ~~~~
    B^{(1)}_{S,y}(r,y_{\pm})
= \mp \frac{4}{3N a_{\pm}^4} \frac{1}{r} \partial_r  \Phi_{\pm}
\label{eq:A_(S,y) at y_pm} \,,
\end{eqnarray}
which are the same as Eq.~(\ref{eq:jun. for A_,y}).

As for the scalar mode, we obtain from Eqs. (\ref{eq:Y^1 zero mode}) and (\ref{eq:xi^y - Phi}) 
\begin{eqnarray}
    Y^{(1)}_0(r,y) + \frac{2H }{\dot\phi_0}\varphi^{(1)}_0(r,y)
    &=& \frac{2}{3} \sum_\sigma   \Phi_\sigma    \,,
\label{eq:comb Y varphi} 
\\
    \bigl( Y^{(1)}_0+2H\hat\xi^y_\pm \bigr)_{y=y_{\pm}}
     &=& \frac{2}{3} \sum _\sigma  \Phi_\sigma  \,.
\label{eq:comb Y 2H xi}
\end{eqnarray}
From Eqs.~(\ref{eq:d varphi = Y^J +...}) and  (\ref{eq:Ein-Y^J}),
\begin{eqnarray}
  ( Y_{,r}^{(1)})^2 + \frac{2\kappa}{3} \bigl(\varphi_{,r}^{(1)}\bigr)^2  
&=& \frac{1}{a^2} \partial_y 
  \Bigl( a^2 Y_{,r}\frac{\varphi _{,r}}{\dot\phi_0} 
  \Bigr)
 + \frac{3 Y_{,r}(\Delta Y)_{,r}}{2\kappa a^2 \dot\phi_0^2} 
 \,,
\label{eq:Y^2+varphi^2}
\\
 \varphi_{,y}^{(1)} 
 &=& - \frac{3 \Delta Y}{2\kappa a^2 \dot\phi_0}
+ \dot\phi_0 Y
+ \frac{\ddot\phi_0 \varphi }{\dot\phi_0} 
\,,
\\
 \varphi_{,yy}^{(1)} 
 &=& 
 - \frac{\Delta \varphi }{a^2}
 + \frac{6H \Delta Y}{ \kappa a^2 \dot\phi_0}
+ 2( \ddot\phi_0 - H \dot\phi_0 )  Y 
+ \left( 
       \frac{2\kappa}{3}  \dot\phi_0^2 
     + \frac{  \stackrel{ \ldots }{\phi_0}  }{\dot\phi_0} 
  \right) \varphi 
\,.
\end{eqnarray}

Integrating by parts, we derive 
\begin{eqnarray}
  \int^{y }_{y_{\sigma}}dy 
   \partial_y (a^2 u_{-\sigma})  
\left( 
     \frac{3}{2 \kappa \dot\phi_0^2} 
 \partial_y  S_\psi 
  \right)
  =  a^2 v_{-\sigma}(y) S_\psi 
  - \int^{y}_{y_{\sigma}}dy   a^2 S_\psi \,.
\label{eq:formula of S_psi}
\end{eqnarray}

For reference, we give the explicit form of Eq. (\ref{eq:def. S_jun}) for $S^{\pm}_{jun}$ as 
\begin{eqnarray}  
 \frac{ \lambda_{\pm}}{2\dot\phi_0} S^{\pm}_{jun} 
  - \frac{ \ddot\phi_0  \varphi^2}{2 \dot\phi_0^3} 
  + \frac{3 \varphi  \Delta  Y }
      {2\kappa a^2 \dot\phi_0^3}
&=&
\frac{\xi^r}{\dot\phi_0} \partial_r (\varphi-\dot\phi_0 \xi^y)
  \pm
\frac{\lambda_\pm}{\dot\phi_0}
\biggl\{
 \frac{\dot\phi_0}{2a^2}(\xi^y_{\pm,r})^2
+ \frac{3 \xi^r_\pm (\Delta Y)_{,r} }{2\kappa a^2 \dot\phi_0} 
+ \frac{3 ({\mathbb S}_Y  + a^2 \dot H  Y^2 )  }{2\kappa a^2 \dot\phi_0}
+ \frac{9 (\Delta Y)^2}{4\kappa^2 a^2 \dot\phi_0^3} 
\nonumber
  \\
  &&
+  
\biggl[
 \frac{3\ddot\phi_0 \Delta Y }{2\kappa a^2 \dot\phi_0^3}
  + \frac{6H \Delta Y}{\kappa a^2 \dot\phi_0^2}
  - \frac{\Delta \varphi}{a^2 \dot \phi_0}
  + \frac{(\varphi-\dot\phi_0 \xi^y)}{2 \dot\phi_0}  \partial_y 
  \Bigl( \frac{\ddot \phi_0}{\dot \phi_0}
  \Bigr)
  + \frac{\xi^y_{\pm,r} }{a^2} \partial_r
\biggr] (\varphi-\dot\phi_0 \xi^y)
\biggr\}
\nonumber
 \\
 &&   
+ (\varphi-\dot\phi_0 \xi^y) 
\left[
 \frac{ 3 \Delta Y }{2\kappa a^2 \dot\phi_0^3}
 -  (\varphi-\dot\phi_0 \xi^y) \Bigl( 
   \frac{\lambda_\pm}{4\dot \phi_0}    
   V'''_{(\pm)} + \frac{\ddot \phi_0}{2\dot\phi_0^3}
   \Bigr)
\right] \,.
\end{eqnarray}

    \subsection{Goldberger-Wise mechanism}
    \label{sec:Integration, GW}

In the text, we assumed that the $y$ integration containing the factor $1/\dot\phi_0^2$ is dominated by the contribution near the negative tension brane. To justify this assumption, we discuss the behavior of the bulk scalar field $\phi_0$.  
For definiteness, we adopt the model proposed by Goldberger and Wise\cite{Goldberger:1999uk}, in which the scalar field potentials are 
\begin{eqnarray}
   V_B(\tilde \varphi)={M^2 \tilde\varphi^2 \over 2},
   \quad
   V_{(\sigma)} (\tilde \varphi) = \gamma_\sigma
     \bigl(\tilde\varphi^2 - h_\sigma^2 \bigr)^2 \,,
\end{eqnarray}
where $h_\sigma \approx \tilde\varphi(y_\sigma) $. $M$ and $\gamma_\sigma$ are the mass and the coupling constant, respectively. 
The scalar field is solved in the weak back reaction approximation as 
\begin{equation}
    \phi_0(y) = B_1 e^{\nu_1 y}+B_2 e^{\nu_2 y},
\label{eq:varphi-solution}
\end{equation}
where  $\nu_1= 2 \ell^{-1}+ \sqrt{4\ell^{-2}+M^2}$, $\nu_2=2\ell^{-1}-\sqrt{4\ell^{-2}+M^2}$, 
and 
\begin{eqnarray}
 && B_1\approx e^{-\nu_1 d}
     \left(\phi_0(y_-)- e^{\nu_2 d} \phi_0(y_+) \right),\cr 
 && B_2\approx \phi_0(y_+) - e^{-\nu_1 d} \phi_0(y_-) .
\end{eqnarray}

According to Ref. \cite{Tanaka:2000er}, the mass squared corresponding to the lowest eigenmode in scalar-type perturbation, $m_S^2$, is estimated by 
\begin{eqnarray}
   N  \int^{d}_{0}dy \frac{1}{\kappa a^4 \dot \phi_0^2}
 &\approx& 
   \frac{1}{3a^2_- m_{S}^2}   \,. 
\end{eqnarray}
Since $\kappa\dot\phi^2_0\lesssim \ell^{-2}$ for the assumption of the weak back reaction to be consistent, 
$m_S^2$ is at most $O(\ell^{-2}a_-^2)$. 
The dominant contribution to this integral comes from the minimum of $(a^4 \dot \phi_0^2)$ at $y=y_c = (\ell \log[\nu_2 B_2/(\nu_1 B_1)])/4\sqrt{1+(M^2\ell^2/4)}$ or $y=d$ when $y_c>d$. 
For convenience, we define a quantity of $O(\ell^{-2})$ by 
$\tilde m_S^2:=a_-^{-2} m_S^2$, absorbing the factor $a_-^2$.  
Following the argument given in Ref.\cite{Tanaka:2000er}, we obtain the estimate 
\begin{equation}
 N\int_{y_+}^{y_-} dy {u_\alpha u_\beta \over\kappa\dot\phi^2_0}
 \approx {1 \over 3 a_\alpha^2 a_\beta^2 \tilde m_S^2}, 
\end{equation}
where $\alpha$ and $\beta$ are $+$ or $-$.  
Here we used the fact that $a^2 u_{\pm}$ is a slowly changing function for $y>y_c$ in the weak back reaction case. 
From this relation, with the aid of inequalities 
$u_-\lesssim a_-^{-2}$ and 
$a_-^2 < u_+ \lesssim 1$, we obtain 
\begin{eqnarray}
  N \int^{d}_{0} dy  
    \frac{u_{\pm} }{\kappa \dot \phi_0^2}
 &\lesssim& 
 \sO \Bigl( \frac{1}{a_-^2 a_{\pm}^2 \tilde m_{S}^2} \Bigr)\,,
\nonumber
\\
  N \int^{d}_{0}dy \frac{u_{-}^i u_+^j}{\kappa \dot \phi_0^2}
 &\lesssim&
  \sO \Bigl(\frac{1}{a_-^{2i }  \tilde m_{S}^2} \Bigr)  
\quad ( \text{for}~~  i,j \geq 0, ~ i+j  \geq 2,).   
\label{eq:approx int}
\end{eqnarray}
The estimate for Eq. (\ref{Eq:YScorrect}) can be obtained by approximating Eq.~(\ref{eq:Y^J - correction}) as  
\begin{eqnarray}
  Y^{(1)}_{S}(r,y) \approx 
  3N u_{\mp}(y)
     \int_{y_+} ^{y_-}  
     \frac{u_{\pm} \Delta Y ^{(1)}_{0}}{\kappa \dot\phi_0^2}  dy' \,, 
\quad( \text{for}~~  y ~~ 
\raisebox{2pt}{$<$}\hspace*{-8pt}
\raisebox{-4pt}{$>$} 
~y_c).   
\end{eqnarray}

    \section{Explicit evaluation of the leading order}
    \label{sec:evaluation}

To derive the leading order of Eq. (\ref{eq:sol bar A^2_[000]}), we give the result of the explicit evaluation of Eq. (\ref{eq:A_[GG0] rew}). We use Eqs. (\ref{eq:A^1 zero mode}), (\ref{eq:Y^1 zero mode}), (\ref{eq:xi^y - Phi}) and the junction condition (\ref{eq:jun. for A_,y}) that is rewritten in terms of the KK mode as 
Eq. (\ref{eq:A_(S,y) at y_pm}).  Keeping the terms of $\sO(r_\star^{-2})$, each term is given as follows: 
\begin{eqnarray}
 - \Delta
   \Bigl[
     {\hat\xi_{ \pm }^y} \bar{A}_{,y}
   + {\hat\xi_{ \pm}^r} \bar{A}_{,r} 
   + \dot H ( {\hat\xi_{ \pm}^y})^2   
\Bigr] 
  = 
   - \Delta \Bigl(
     {\hat\xi_{ \pm }^y} \bar{A}_{,y}
 + 2 \Phi_{\pm,r} \hat\xi_{\pm}^r
 + \frac{ \dot H}{9N^2 a_{\pm}^4} \Phi_\pm ^2
   \Bigr) \,,
\end{eqnarray}
\begin{eqnarray}
   \pm 2N  a_{\pm}^4 S_{\Sigma}^{\pm} 
 &=&  
   \frac{16}{9} \int dr 
\Bigl\{
 \Bigl(
   \frac{3\Phi_{\pm,r}}{r^5} - \frac{\Delta \Phi_\pm}{r^4} 
 \Bigr)
  \int r^2 \Phi_\pm dr 
  - \frac{ \Phi_\pm \Phi_{\pm,r} }{ r^2} 
  + \frac{ \Phi_\pm \Delta \Phi }{3r}  
  + \frac{4(\Phi_{\pm,r})^2}{3r} 
  -  \Phi_{\pm,r} \Delta \Phi_\pm 
\nonumber
\\
&& 
  + \frac{ \Phi_{\pm,r} (\Delta \Phi_{\pm})_{,r} }{4} 
 \Bigr\} 
- \frac{4}{3} (\Phi_{\pm,r})^2 
- 4 \Phi_{\pm } \Delta \Phi_{\pm} 
- \frac{8}{3} \xi^r  (\Delta \Phi_{\pm})_{,r}
\nonumber 
\\
&&
\pm 
    \frac{2H}{9N a_{\pm}^2} 
\biggl\{
  (\Phi_{\pm,r})^2 - 26  \Phi_\pm \Delta \Phi 
+
 2 \int dr 
\Bigl[
   \frac{4}{r} (\Phi_{\pm,r})^2
  + \Phi_{\pm } (\Delta \Phi_{\pm})_{,r}
 \Bigr] 
\biggr\} \,,
\end{eqnarray}

\begin{eqnarray}
   \pm 2N  a_{\pm}^4 S_{\xi}^{\pm} 
 &=&  
- \frac{2}{3}  
   \hat\xi^r (\Delta \Phi_\pm)_{,r}  
 \mp \frac{4H}{9N  a_{\pm}^2} 
 \int dr \Bigl[
      \Phi_\pm (\Delta \Phi_\pm)_{,r}
   + \frac{6}{r} (\Phi_{\pm,r} )^2
 \Bigr]
 +  \frac{64}{9}  \int dr 
\Bigl[
-  \frac{\Phi_\pm \Phi_{\pm,r} }{r^2} 
+  \frac{\Phi_\pm \Delta \Phi_\pm }{3 r}
\nonumber
\\
&&
-  \frac{ (\Phi_{\pm,r})^2}{6 r} 
+  \frac{ \Phi_{\pm,r} \Delta \Phi_\pm }{2} 
+
 \Bigl( 
   \frac{3 \Phi_{\pm,r}}{r^5} 
    -  \frac{ \Delta \Phi_\pm }{r^4} 
    +  \frac{ (\Delta \Phi_{\pm})_{,r} }{4 r^3} 
\Bigr)
    \int r^2 \Phi_\pm dr
\Bigr] \,,
\end{eqnarray}

\begin{eqnarray}
  2N \int dy a^2 S_{A\pm} 
&=&  \frac{64}{9} \int dr 
\biggl[
- \frac{5  \Phi_\pm ^2}{3 r^3} 
+  \frac{20 \Phi_\pm \Phi_{\pm,r} }{3 r^2} 
-  \frac{ 3 (\Phi_{\pm,r})^2}{2r}
-  \frac{2 \Phi_\pm \Delta \Phi_\pm}{r}
+  \int r^2 \Phi_\pm dr 
\Bigl(
- \frac{ 15 \int r^2  \Phi_\pm  dr }{ r^9} 
+ \frac{ 10  \Phi_\pm }{ r^6}
\nonumber \\
&&
-  \frac{20  \Phi_{\pm,r}}{r^5} 
+ \frac{ 6  \Delta \Phi_{\pm}}{r^4} 
-  \frac{ (\Delta \Phi_{\pm})_{,r} }{r^3}
\Bigr)
\biggr]
+  \frac{16}{3} \Bigl(1 \pm \frac{H } {Na^2}\Bigr)
\Bigl( 
   \int \frac{ (\Phi_{\pm,r})^2 }{6r} dr
- \Phi_\pm \Delta \Phi_\pm
\Bigr) 
  + \sO \left( \frac{1}{r_\star^4} \right)   
   \,,
\end{eqnarray}
 
\begin{eqnarray}
  2N \int dy a^2  S_{\psi\pm} 
&=&   \frac{64}{9} \int dr 
\biggl\{
  \frac{ 3 \Phi_{\pm,r} \Delta \Phi_\pm}{4}  
- \frac{3(\Phi_{\pm,r})^2}{4r}
- \frac{ \Phi_\pm \Delta \Phi_\pm }{4r }
+ \frac{17 \Phi_\pm \Phi_{\pm,r} }{12r^2}
- \frac{5 \Phi_\pm ^2}{3r^3} 
\nonumber 
\\
&&
+ \Bigl[ 
  \frac{ 3 \Delta \Phi_\pm }{4 r^4}
 - \frac{17  \Phi_{\pm,r} }{4 r^5} 
 + \frac{10  \Phi_\pm }{r^6}  
 - \frac{15 \int r^2 \Phi_\pm dr }{r^{9}}
\Bigr]
  \int r^2 \Phi_\pm  dr 
\biggr\} 
\nonumber
\\
&&
+ \Bigl(1 \pm \frac{H } {Na^2}\Bigr)
\Bigl(  
  \frac{2}{3} (\Phi_{\pm,r})^2 
- \frac{16}{9} \int \frac{ dr }{r}(\Phi_{\pm,r} )^2 
\Bigr) 
  + \sO \left( \frac{1}{r_\star^4} \right)  
\,,
\end{eqnarray}
\begin{eqnarray}
 2N \int^{y_{-}} _{y_{+} }dy' 
   a^2 v_{\pm} \Delta  {\mathbb S}_{\varphi\pm}
&& 
 = \Delta( \Phi_\pm^2  )
\biggl[ 
      \frac{2}{9} \Bigl( 1 \pm \frac{H  }{a_{\pm}^2 N} \Bigr)
    + \frac{\dot H}{ 9 N^2 a_{\pm}^4 }   
    - \frac{4N}{9} \int dy a^2 u_\pm^3  
\biggr] 
+ \sO \left( \frac{ 1}{r_\star^4} \right) \,,
\end{eqnarray}

\begin{eqnarray}
3N  \int^{y_-}_{y_+}dy' 
     \frac{u_{\pm}  \Delta {\mathbb S}_{Y\pm}} {\kappa \dot \phi_0^2} 
&=&
    \frac{4N \Delta (\Phi_\pm^2) }{9} \int dy a^2 u_\pm^3
  + \sO \left( \frac{ 1}{r_\star^4} \right) \,.
\end{eqnarray}
Substituting these results into Eq. (\ref{eq:A_[GG0] rew}), we obtain  the leading term of Eq. (\ref{eq:sol bar A^2_[000]}).


\end{document}